\documentclass[lettersize,journal]{IEEEtran}
\usepackage{graphicx}
\usepackage{cite}
\usepackage{picinpar}
\usepackage{amsmath,amssymb,amsfonts}
\usepackage{url}
\usepackage{flushend}
\usepackage[latin1]{inputenc}
\usepackage{colortbl}
\usepackage{soul}
\usepackage{multirow}
\usepackage{pifont}
\usepackage{color}
\usepackage{alltt}
\usepackage[hidelinks]{hyperref}
\usepackage{enumerate}
\usepackage{siunitx}
\usepackage{breakurl}
\usepackage{epstopdf}
\usepackage{pbox}
\usepackage{bm}

\usepackage{multirow}
\usepackage{tabularx}
\usepackage{subfigure}  
\usepackage{booktabs}
\usepackage{stfloats}
\usepackage{enumerate}
\usepackage{algpseudocode}
\usepackage[linesnumbered,ruled]{algorithm2e}
\usepackage{amsmath} 
\usepackage{tikz}
\usepackage{tikz}
\usepackage{verbatim}

\usepackage{bm}

\newcommand{\Rmnum}[1]{\uppercase\expandafter{\romannumeral #1}}  

\makeatletter
\newcommand\figcaption{\def\@captype{figure}\caption}
\newcommand\tabcaption{\def\@captype{table}\caption}
\makeatother

\usepackage{soul}
\newcommand{\RNum}[1]{\uppercase\expandafter{\romannumeral #1\relax}}

\usepackage{nomencl}
\makenomenclature

\begin{document}
\title{Unsupervised Attention-Based Multi-Source Domain Adaptation Framework for Drift Compensation in Electronic Nose Systems}

\author{

	Wenwen Zhang, \emph{Member IEEE},
	Shuhao Hu, 
	Zhengyuan Zhang,
	Yuanjin Zheng, \emph{Senior Member IEEE},
	Qi Jie Wang, \emph{Senior Member IEEE},
	Zhiping Lin, \emph{Senior Member IEEE}

	\thanks{This work was supported in part by the  National Natural Science Foundation of China under Grant 62203307, National Medical Research Council (NMRC) MOH-000927, and A*STAR grants (A2090b0144 and R22I0IR122) (\textit{Corresponding authors: Zhiping Lin.})  \par		
	Wenwen Zhang, Zhengyuan Zhang, Yuanjin Zheng, Qi Jie Wang and Zhiping Lin are with the School of Electrical and Electronics Engineering, Nanyang Technological University, Singapore 639798, Singapore (e-mail: wenwen.zhang@ntu.edu.sg; zhengyuan.zhang@ntu.edu.sg; YJZHENG@ntu.edu.sg; qjwang@ntu.edu.sg; ezplin@ntu.edu.sg)\par
	Shuhao Hu is with the School of Sino-German College of Tongji University, Shanghai 201804, China (e-mail: TJhushuhao@tongji.edu.cn)    
}
}

\maketitle
	
\begin{abstract}
Continuous, long-term monitoring of hazardous, noxious, explosive, and flammable gases in industrial environments using electronic nose (E-nose) systems faces the significant challenge of reduced gas identification accuracy due to time-varying drift in gas sensors. To address this issue, we propose a novel unsupervised attention-based multi-source domain shared-private feature fusion adaptation (AMDS-PFFA) framework for gas identification with drift compensation in E-nose systems. The AMDS-PFFA model effectively leverages labeled data from multiple source domains collected during the initial stage to accurately identify gases in unlabeled gas sensor array drift signals from the target domain. To validate the model's effectiveness, extensive experimental evaluations were conducted using both the University of California, Irvine (UCI) standard drift gas dataset, collected over 36 months, and drift signal data from our self-developed E-nose system, spanning 30 months. Compared to recent drift compensation methods, the AMDS-PFFA model achieves the highest average gas recognition accuracy with strong convergence, attaining 83.20\% on the UCI dataset and 93.96\% on data from our self-developed E-nose system across all target domain batches. These results demonstrate the superior performance of the AMDS-PFFA model in gas identification with drift compensation, significantly outperforming existing methods. 
\end{abstract}


\markboth{IEEE TRANSACTIONS ON xxx}
{}

\printnomenclature

\section{Introduction}
\IEEEPARstart{A}{n} electronic nose (E-nose) system mimics the mammalian olfactory system to identify components in mixed odors \cite{b1, b2, b3}. However, gas sensor drift poses a significant challenge, undermining long-term identification accuracy \cite{b20}. Efforts to address this focus on developing advanced gas sensors with better resistance to aging, poisoning, and environmental changes \cite{b22}. Despite these advancements, implementing them in practical applications remains time-consuming and costly.\par

The prevailing practice at most industrial sites is to procure established commercial gas sensors directly for on-site monitoring.  Consequently, more  efforts are directed towards developing the novel gas identification models,  incorporating drift compensation techniques to process signals from gas sensor arrays \cite{r22}.  For instance, in light of the remarkable accomplishments of domain adaptation (DA) within the field of transfer learning, some researchers are now channeling their efforts into the development of novel DA  models designed to effectively address the challenge posed by gas sensor drift. For example, to investigate the cross-domain learning capability of extreme learning machines (ELM),   Zhang \textit{et al} proposed the domain adaptation extreme learning machine (DAELM) method \cite{b36} to develop robust classifiers for compensating drift and identifying gases in E-nose systems. This is achieved by utilizing a limited quantity of labeled data from the target domain and a single source domain. However, in real-world scenarios,  annotated gas sensor data can originate from multiple source domains. Moreover, the label for the target domain is frequently unspecified or unknown. \par
Researchers have developed novel models that leverage labeled data from multiple source domains to address tasks such as classification and fault diagnosis with unlabeled target domain data \cite{b30, b31, b32}. However, most existing multi-source domain models focus on extracting invariant features across domains \cite{c9, c10} while neglecting unique source-target domain features. To improve drift suppression in E-nose systems, our study introduces the unsupervised attention-based multi-source domain shared-private feature fusion adaptation (AMDS-PFFA) model. This model leverages labeled data from multiple source domains and utilizes shared-private feature fusion to effectively counteract gas sensor drift while minimizing discrepancies among classifiers.\par
We verified our approach using both the gas sensor array drift dataset from the Biocircuits Institute at the University of California, Irvine (UCI) \cite{b34} and data collected over 30 months with a customized E-nose system designed in our laboratory.
The main contributions of our work are summarized as follows:\par 
\begin{enumerate}[ 
	\setlength{\parindent}{2em}  1)]
    \item Drawing inspiration from the shortcut architecture of Residual Networks (ResNet), our study presents a novel multi-source domain shared-private feature fusion framework that integrates unique private features from each source-target pair. This approach overcomes the common limitation of existing multi-source domain adaptation models, which focus only on domain-invariant features, thereby significantly improving gas recognition accuracy and drift compensation in the target domain.\par
    \item  The AMDS-PFFA model is an unsupervised approach that leverages labeled data from multiple source domains to perform gas identification and drift compensation on unlabeled target domain signals, effectively mitigating drift without requiring additional labels. It enhances accuracy by refining local maximum mean discrepancy (LMMD) for feature alignment and harmonizing probability prediction outputs across classifiers. \par 
	\item Gas sensor drift is unpredictable, making it challenging to design methods that consistently maintain high gas recognition accuracy over time.  The AMDS-PFFA model achieves the highest average accuracy for drift compensation across all UCI Chemosignals Lab data batches and shows superior convergence, outperforming recent models. Furthermore, its practical application has been successfully validated within our self-developed E-nose system, where it also achieved the highest gas recognition accuracy with strong convergence. \par
\end{enumerate}

The subsequent sections of this article is organized as follows. Section \Rmnum{2} provides an overview of related works. In  Section \Rmnum{3}, we elucidate the framework of the unsupervised AMDS-PFFA model. The experimental validation, utilizing the UCI Chemosignals Lab drift dataset, is introduced in Section \Rmnum{4}. Section \Rmnum{5} presents  our self-developed E-nose system and outlines experimental validation using drift data obtained from self-developed E-nose. Finally, Section \Rmnum{6} summarizes the key conclusions drawn from this study.  
\section{related-works}
\subsection{Unsupervised Multi-source domain adaptation}
The unsupervised multi-source domain adaptation (MSDA) method addresses the challenge of learning a classifier for an unlabeled target domain by leveraging knowledge from  multiple labeled source domains, Currently,  a variety of MSDA models are being proposed  to improve the target domain task identification accuacy \cite{c8}. For example, Chai \textit{et al.} introduce a multisource-refined transfer network to address fault diagnosis issues under both domain and category inconsistencies \cite{b40}. Chen \textit{et al} have devised a multi-source weighted deep transfer network for open-set fault diagnosis in rotary machinery \cite{b32}. This network leverages an open-set adversarial training module and an adaptive weighting learning module, collaboratively constructed to acquire domain-invariant features that are applicable to different  machine health conditions.        

\subsection{Gas sensor drift compensation}
From a data distribution perspective, drifted data is treated as the target domain, while un-drifted data is considered the source domain \cite{r24, r25, r26}. Current research focuses on finding mappings that align the projections of both clusters in a common space, aiming for distribution consistency. For example, Zhang \textit{et al.} proposed domain regularized component analysis (DRCA) to enhance distribution alignment and drift adaptation in principal component subspaces \cite{b25}. They also developed an unsupervised feature adaptation (UFA)-based transfer learning method to improve drift tolerance in E-noses \cite{r27}. Building on this, Yi \textit{et al.} introduced local discriminant subspace projection (LDSP)  \cite{b26}, extending DRCA by using label information to better separate samples in the latent subspace. Yan \textit{et al.} proposed domain correction latent subspace learning (DCLSL) \cite{b27}, which aligns statistical distributions before and after drift while leveraging geometric structure and domain discriminative information . Wang \textit{et al.} tackled baseline drift in optical E-noses using independent component analysis \cite{r28}. Unlike conventional subspace methods, Zhang \textit{et al.} framed gas sensor drift as a cross-domain recognition problem and introduced cross-domain discriminative subspace learning (CDSL) \cite{b28} to address odor recognition challenges . Yi \textit{et al.} further advanced subspace learning with a novel discriminative domain adaptation neighborhood preserving (DANP) method \cite{b29}, bypassing assumptions of specific data distributions like Gaussian. Some researchers have turned to deep learning models, such as Feng \textit{et al.} introduced an augmented convolutional neural network (ACNN) \cite{r23} to continuously address gas sensor drift with high accuracy over extended periods. \par
\section{unsupervised AMDS-PFFA model for Gas IDentification with drift compensation }  
\subsection{Unsupervised AMDS-PFFA model framework}
\begin{figure}[htbp]\centering
	\includegraphics[width=8.5cm]{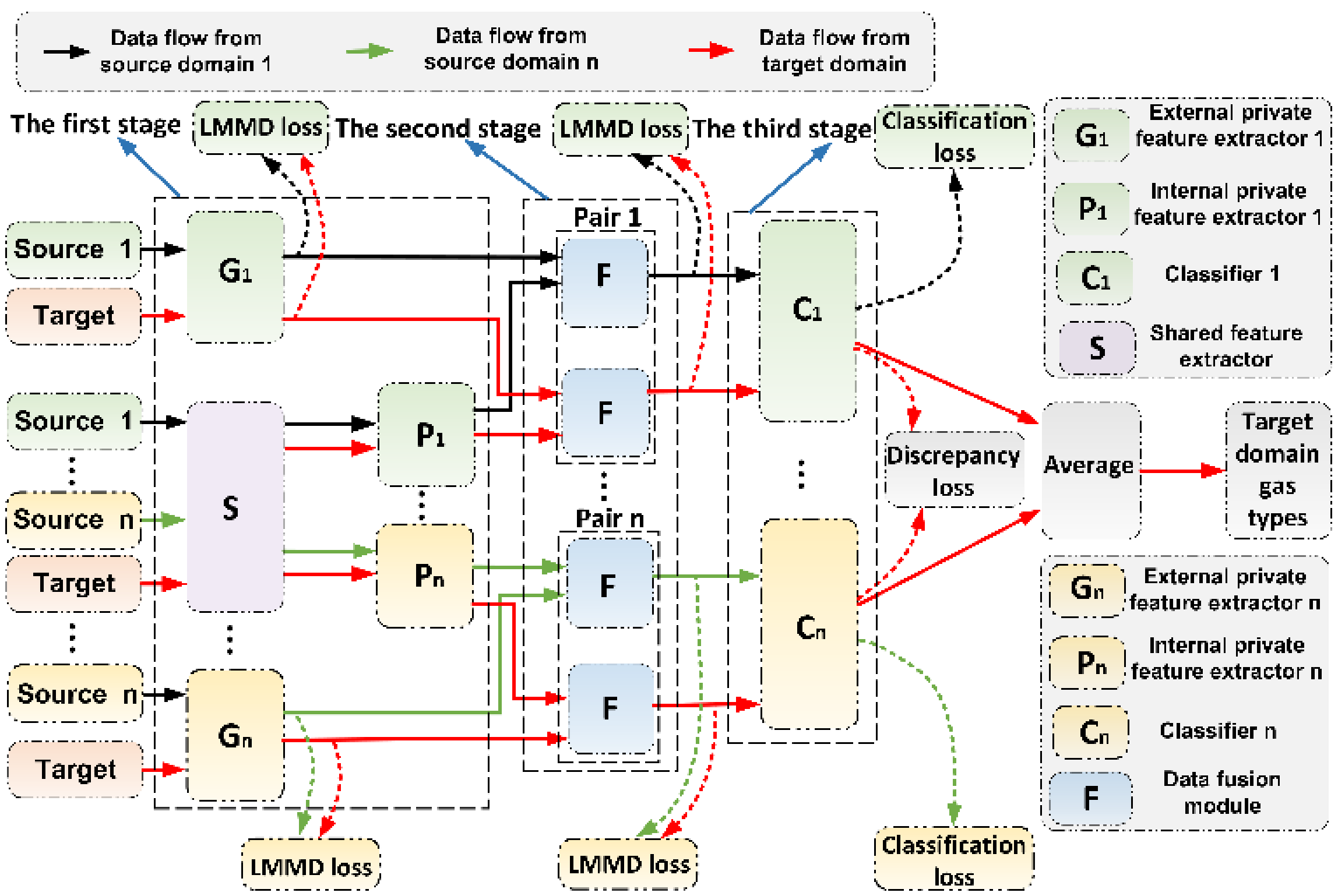}
     \caption{Overview of the proposed unsupervised AMDS-PFFA model framework for drift compensation.}
	\label{framework}
\end{figure}

To effectively utilize labeled gas data from multiple source domains and enhance gas identification accuracy with drift compensation in unlabeled, drifted target domains, an unsupervised attention-based AMDS-PFFA  gas identification model framework are proposed for drift counteraction in this work. The overview of the proposed unsupervised AMDS-PFFA model framework for drift compensation is  depicted in Fig. \ref{framework}. It consists of three stages. The detailed description of the model framework and the specific implementation process are as follows:  \par
\subsubsection{The first stage - feature extractors}as illustrated in Fig. \ref{framework}, consider a set of \( n \) labeled source domains, denoted as \(\{(\mathcal{D}_{s_j}, \mathcal{Y}_{s_j})\}_{j=1}^{N_s}\), where \( N_s \) denotes the number of source domains, corresponding to the variable \( n \) as shown in Fig. \ref{framework}. For each source domain \( \mathcal{D}_{s_j} \), we have gas sensor array signal features \(\{\mathbf{x}_i^{s_j}\}_{i=1}^{|\mathcal{D}_{s_j}|}\), and corresponding gas type labels \(\mathcal{Y}_{s_j} = \{\mathbf{y}_i^{s_j}\}_{i=1}^{|\mathcal{D}_{s_j}|}\), where \(\mathbf{y}_i^{s_j} \in \mathbb{R}^{K}\) with \( K \) representing the number of gas types. Each source domain exhibits a distinct data distribution \(\{p_{s_j}\}_{j=1}^{N_s}\). Additionally, there is a target domain with unlabeled gas data, denoted as \(\mathcal{D}_{t}\), where \(\mathcal{D}_{t} = \{\mathbf{x}_{i}^{t}\}_{i=1}^{|\mathcal{D}_t|}\). The data distribution of the target domain is \(p_{t}\) and differs from the source domain distributions, i.e., \(p_{t} \neq \{p_{s_j}\}_{j=1}^{N_s}\). In the first stage, the model includes \( n \) external private feature extractors for each source-target domain pair and one common feature extractor. This common feature extractor is cascaded with \( n \) internal private feature extractors, one for each source-target domain pair. The external private feature extractors are denoted as  \({G}_1, \ldots, {G}_n\), the common feature extractor as \( S \), and the internal private feature extractors as \({P}_1, \ldots, {P}_n\).\par

\textit{(a) Shared feature extractor cascaded with $n$ internal private feature extractors:}  
the shared feature extractor $ S$ maps all domains into a unified feature space to extract common features across them. Given gas sensor array signal features $\mathbf{x}^{s_j}$ from the source domain $(\mathcal{D}_{s_j}, \mathcal{Y}_{s_j})$ and $\mathbf{x}^{t}$ from the target domain $\mathcal{D}_t$, the cascaded internal private feature extractors process the common features $S(\mathbf{x}^{s_j})$ and $S(\mathbf{x}^{t})$. Each of the $n$ internal private feature extractors then maps its corresponding source-target domain pair into distinct private feature spaces. The final output features from the $j$-th internal private feature extractor are $\mathcal{F}_{s_j}^{in} = P_j(S(\mathcal{D}_{s_j}))$ and $\mathcal{F}_{t_j}^{in} = P_j(S(\mathcal{D}_{t}))$, respectively.\par
\textit{(b) External private feature extractors:}     
as illustrated in Fig. 1, there are $n$ source-target domain pairs and $n$ corresponding external private feature extractors, labeled $ G_1, \ldots, G_n$. Unlike shared extractors, these do not share weights and are designed to directly capture the private features of each source-target domain pair. Given gas sensor array signal features $\mathbf{x}^{s_j}$ from source domain $(\mathcal{D}_{s_j}, \mathcal{Y}_{s_j})$ and $\mathbf{x}^{t}$ from target domain $\mathcal{D}_t$, the external private feature extractor $G_j$ extracts features $G_j(\mathbf{x}^{s_j})$ and $G_j(\mathbf{x}^t)$ for the source and target domains, respectively. The final output features from the external extractor are expressed as $\mathcal{F}_{s_j}^{ex} = G_j(\mathcal{D}_{s_j})$ and $\mathcal{F}_{t_j}^{ex} = G_j(\mathcal{D}_{t})$.\par
\textit{(c) Local maximum mean discrepancy LMMD loss:} to capture fine-grained information for each gas type across corresponding groups in the source domain $\mathcal{D}_{s_j}$ and the target domain $\mathcal{D}_t$, we divide the gas data in the source domain into $K$ distinct subdomains, denoted as $\mathcal{D}_{s_j}^{k}$ (where $k=1, \ldots, K$), with each subdomain representing one of the $K$ gas types. Similarly, the target domain is segmented into $K$ subdomains, denoted as $\mathcal{D}_{t}^{k}$. To align the feature distributions of each gas type within these subdomains, we employ the local maximum mean discrepancy (LMMD) metric 
\cite{b35}. The LMMD between the $j$-th source domain $\mathcal{D}_{s_j}$ and the target domain $\mathcal{D}_t$ in the Reproducing Kernel Hilbert Space (RKHS) is estimated as follows:
\vspace{-10pt} 
\begin{equation}\label{f1}
\small
\begin{aligned}
\hat{D}_{\mathcal{H}}(p_{s_j}, {p}_{t})\!=\!\!\frac{1}{K}\!\sum_{k=1}^{K}\!\left\|\!\sum_{\mathbf{x}_{m}^{s_j}\in{\mathcal{D}_{s_j}}}\!\!\!\!\!\!\!\lambda_m^{s_{j}k}\phi(\mathbf{x}_{m}^{s_j})-\!\!\!\!\sum_{\!\mathbf{x}_{n}^{t}\in\mathcal{D}_t}\!\lambda_n^{t{k}}\phi(\mathbf{x}_{n}^{t})\!\right\|_{\mathcal{H}}^{2}
\end{aligned}
\end{equation}
where $\phi(\cdot)$ denotes a feature mapping function that projects the original samples into the RKHS, while $\lambda_m^{s_jk}$ and $\lambda_n^{tk}$ are the probability coefficients corresponding to the source domain sample $\mathbf{x}_m^{s_j}$ and the target domain sample $\mathbf{x}_n^{t}$, respectively, both of which belong to gas type $k$. These coefficients satisfy $\sum\limits_{m=1}^{M} \lambda_m^{s_jk} = 1$ and $\sum\limits_{n=1}^{N} \lambda_n^{tk} = 1$, where $M$ and $N$ denote the number of gas sensor array signal samples in the $j$-th source and target domains, respectively. The weighted sums for gas type $k$ are expressed as $\sum_{\mathbf{x}_m \in \mathcal{D}_{s_j}} \lambda_m^{s_jk} \phi(\mathbf{x}_m^{s_j})$ and $\sum_{\mathbf{x}_n \in \mathcal{D}_t} \lambda_n^{tk} \phi(\mathbf{x}_n^{t})$. The coefficients $\lambda_m^{s_jk}$ and $\lambda_n^{tk}$ are computed as follows:
\begin{equation}\label{f2}
\small
\begin{aligned}
\lambda_m^{s_jk}=\frac{y_{mk}^{s_j}}{\sum_{(\mathbf{x}_n, \mathbf{y}_n)\in \mathcal{D}_{s_j}}y_{nk}^{s_j}},\quad \lambda_n^{tk}=\frac{y_{nk}^{t}}{\sum_{(\mathbf{x}_m, \mathbf{y}_m)\in \mathcal{D}_{t}}y_{mk}^{t}} 
\end{aligned}
\end{equation}
for the labeled samples in the source domain, the known label $\mathbf{y}_m^{s_j}$ is directly utilized as a one-hot vector to calculate $\lambda_m^{s_jk}$, where $y_{mk}^{s_j}$ denotes the $k$-th entry of $\mathbf{y}_m^{s_j}$. Since the AMDS-PFFA framework is unsupervised, direct calculation of Eq. (\ref{f1}) for the unlabeled target domain $\mathcal{D}_t$ is not feasible due to the unknown $y_{nk}^t$. Instead, we employ the output $\mathbf{\bar{y}}^{t}_n = f_j(\mathbf{x}_n^{t})$ from the $j$-th classifier $C_j$, which provides the probability of assigning $\mathbf{x}_n^{t}$ to each of the $K$ classes. The soft prediction $\bar{y}^t_{nk}$ is then utilized to calculate $\lambda_n^{tk}$, where $f_j(\cdot)$ is the mapping function of $\mathbf{x}_n^t$ to the gas type predicted by classifier $C_j$. This allows the utilize of Eq. (\ref{f1}) to calculate the LMMD value. To integrate this with the deep learning network in the AMDS-PFFA framework, the activations generated by gas sensor array signal samples at the $l$-th layer of the network are denoted as $\{\bm{b}_m^{s_jl}\}_{m=1}^M$ and $\{\bm{b}_n^{tl}\}_{n=1}^N$ for the $j$-th source and target domains, respectively. Eq. (\ref{f1}) is then reformulated as:
\begin{equation}\label{f3}
\small
\begin{aligned}
&\hat{D}_{l}(p_{s_j}, {p}_{t})=\frac{1}{K}\!\sum_{k=1}^{K}\Bigg[\sum_{m=1}^M\sum_{n=1}^M\lambda_m^{s_jk}\lambda_n^{s_jk}k(\bm{b}_m^{s_jl},  \bm{b}_n^{s_jl})+\\&\sum_{m=1}^N\sum_{n=1}^N\lambda_m^{tk}\lambda_n^{tk}k(\bm{b}_m^{tl},  \bm{b}_n^{tl})-2\sum_{m=1}^M\sum_{n=1}^N\lambda_m^{s_jk}\lambda_n^{tk}k(\bm{b}_m^{s_jl},  \bm{b}_n^{tl})\Bigg]
\end{aligned}
\end{equation}
where the kernel $k$ is defined as $k(\mathbf{x}^{s_j}, \mathbf{x}^t)=\langle \phi(\mathbf{x}^{s_j}),  \phi(\mathbf{x}^t)\rangle$, with $\langle\cdot,\cdot\rangle$ denoting the inner product of vectors. $\bm{b}_m^{s_jl}$ and $\bm{b}_n^{tl}$ denote the $m$-th and $n$-th sensor array signal samples in the $j$-th source domain and the target domain, respectively, at the $l$-th layer of network activation. Eq. (\ref{f3}) is employed to describe the distributions of subdomains within the same category across both source and target domains. The LMMD loss for the $N_s$ external private feature extractors, denoted as $\mathcal{L}_{lmmd_1}$, is expressed as follows:
\begin{equation}\label{f4}
\small
\begin{aligned}
\mathcal{L}_{\text{lmmd}_1} = \frac{1}{N_s} \sum_{j=1}^{N_s} \hat{D}_{l}\left(\mathcal{F}_{s_j}^{ex}, \mathcal{F}_{t_j}^{ex}\right)
\end{aligned}
\end{equation}
where $\mathcal{F}_{s_j}^{ex}$ and $\mathcal{F}_{t_j}^{ex}$ denote the output features from the external private feature extractor for the source and target domains, respectively.
\subsubsection{The second stage - feature fusion module}    
\begin{figure*}[htbp]\centering
	\includegraphics[width=17.6cm]{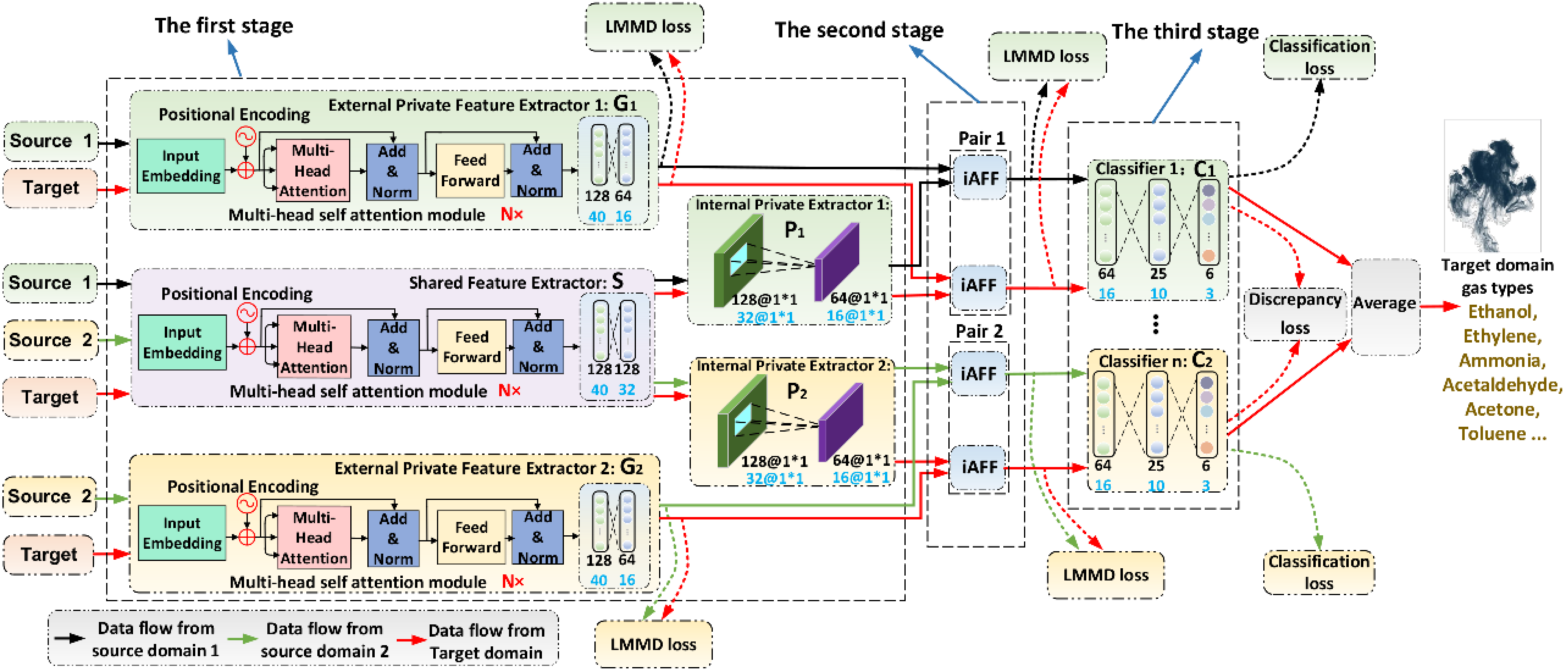}
	\caption{The architecture of unsupervised AMDS-PFFA model for drift compensation.}
	\label{model}
\end{figure*} 

in the first stage, mapping all domains into a shared feature space may lead to the loss of important private features unique to each source-target domain pair. To address this, we draw inspiration from the shortcut connections in ResNet. In the second stage, we propose utilizing an attention-based feature fusion module to focus on preserving the private feature information specific to each source-target domain pair. This module effectively integrates the private feature outputs from both internal and external private feature extractors. The fused features of the $j$-th source domain, denoted as $\mathcal{\bar{D}}_{s_j}$, and the corresponding target domain, denoted as $\mathcal{\bar{D}}_{t_j}$, are formulated as follows:
\begin{equation}\label{f5}
\small
\begin{aligned}
\mathcal{\bar{D}}_{s_j} = F\left(\mathcal{ {F}}_{s_j}^{ex}, \mathcal{ {F}}_{s_j}^{in}\right), \quad  
\mathcal{\bar{D}}_{t_j} = F\left(\mathcal{ {F}}_{t_j}^{ex}, \mathcal{ {F}}_{t_j}^{in}\right)
\end{aligned}
\end{equation}
where $F(\cdot; \cdot)$ denotes the feature fusion function, while $\mathcal{F}_{s_j}^{in}$ and $\mathcal{F}_{t_j}^{in}$ denote the output features from the internal feature extractors for the source and target domains, respectively. The divergence in data distribution between the fused features of the source and target domains is further reduced through the minimization of the $\mathcal{L}_{lmmd_2}$ loss, expressed as follows:
\begin{equation}\label{f6}
\small
\begin{aligned}
\mathcal{L}_{lmmd_2} = \frac{1}{N_s} \sum_{j=1}^{N_s}\hat{D}_{l}(\mathcal{\bar D}_{s_j}, \mathcal{\bar D}_{t_j})
\end{aligned}
\end{equation}
\subsubsection{The third stage - gas type classifiers}  
this stage comprises $N_s$ gas category classifiers, denoted as $\{{C_j}\}_{j=1}^{N_s}$. Each classifier $C_j$ is a softmax classifier that receives fused source domain features from the $j$-th external and internal private feature extractors in the second stage. Each gas classifier employs a cross-entropy loss function. The total cross-entropy loss across the $N_s$ gas category classifiers is defined as:  \par
\begin{equation}\label{f7}
\small
\begin{aligned}
\mathcal{L}_{cross} \! = \!\sum_{j=1}^{N_s} \mathbf{E}_{\mathbf{x} \sim \mathcal{D}_{s_j}} J\left(C_j\left(F\left(P_j(S(\mathbf{x}_{i}^{s_j})),\! G_j(\mathbf{x}_{i}^{s_j})\right)\right), \mathbf{y}_i^{s_j}\right)
\end{aligned}
\small
\end{equation}
where $J(\cdot,\cdot)$ denotes cross-entropy calculation function, and $\mathbf{E}_{\mathbf{x} \sim \mathcal{D}_{s_j}}$ denotes the expected value taken over samples $\mathbf{x}$ drawn from the $j$-th source domain $\mathcal{D}_{s_j}$, reflecting the average cross-entropy loss for these samples, and $\mathbf{y}_i^{s_j}$ denotes the true label for the sample $\mathbf{x}_i^{s_j}$ within the $j$-th source domain.  Since the gas category classifiers are trained on different source domains, their predictions on target samples may yield varying gas identification results for the same target domain. To minimize this divergence, the difference in probability prediction outputs across all classifiers is used as the classifier difference loss. The classifier difference loss between all classifiers is expressed as follows:\par
\vspace{-10pt}
\begin{equation}\label{f9}
\footnotesize
\begin{aligned}
&\mathcal{L}_{diff} = \frac{2}{N_s(N_s-1)} \sum_{j=1}^{N_s-1} \sum_{i=j+1}^{N_s} \mathbf{E}_{\mathbf{x} \sim \mathcal{D}_t} \\ & \left[  \left|   C_i \left( F \left( P_i \left( S(\mathbf{x}_k) \right), G_i(\mathbf{x}_k) \right) \right)  - C_j \left( F \left( P_j \left( S(\mathbf{x}_k) \right), G_j(\mathbf{x}_k) \right) \right) \right| \right]
\end{aligned}
\end{equation}  
where \(\mathbf{E}_{\mathbf{x} \sim \mathcal{D}_t}\) denotes the expected value of the function calculated over samples $\mathbf{x}$ drawn from the target domain $\mathcal {D}_t$, reflecting the average discrepancy between the predictions made by different classifiers for target domain samples. By minimizing Eq. (\ref{f9}), the probability outputs of all classifiers are encouraged to converge, ensuring consistency among their predictions. Ultimately, the predicted gas type for each target domain sample is determined by averaging the outputs of all gas classifiers. \par
In summary, the loss function of the proposed unsupervised AMDS-PFFA framework consists of four components: $\mathcal{L}_{lmmd_1}$, $\mathcal{L}_{lmmd_2}$, $\mathcal{L}_{cross}$, and $\mathcal{L}_{diff}$. Minimizing $\mathcal{L}_{lmmd_1}$ and $\mathcal{L}_{lmmd_2}$ aligns the distribution of gas features within their respective subdomains, enabling the learning of domain-invariant features and private features specific to each source-target domain pair. Minimizing $\mathcal{L}_{cross}$ allows the classifier to accurately identify gas types within the source domain, while minimizing $\mathcal{L}_{diff}$ ensures consistent convergence in the probability outputs across all classifiers. The total loss function, denoted as $\mathcal{L}_{total}$, is therefore formulated as follows: \par
\begin{equation}\label{f10}
\small
\begin{aligned}
\mathcal{L}_{total}=\mathcal{L}_{cross}+\lambda(\mathcal{L}_{lmmd_1}+\mathcal{L}_{lmmd_2}++\mathcal{L}_{diff})
\end{aligned}
\end{equation} 
where $\lambda$ is a configurable coefficient defined as $\lambda = \frac{2}{1 + e^{-\alpha (\frac{\text{epoch}}{\text{epochs}})}}$, and  $\alpha$ can be adjusted based on the experimental data. The adjustment coefficient $\alpha$ employed in our two real-world data scenarios is provided in Tables \ref{para1} and \ref{para2}, respectively. During model training, the total loss $\mathcal{L}_{total}$ is minimized,  as shown in Eq. (\ref{f10}), to facilitate gas identification with drift compensation in the target domain. The process of optimizing AMDS-PFFA model parameters via this total loss function is detailed in Algorithm 1. \par
\begin{algorithm}[htbp]
	\scriptsize
	\caption{Unsupervised AMDS-PFFA Framework}
	\LinesNumbered
	\KwIn{Source domains: $\{(\mathcal{D}_{s_j}, \mathcal{Y}_{s_j})\}_{j=1}^{N_s}$, target domain: $\mathcal{D}_t$, learning rate: $l$, batch size: $b$, weight decay: $w$, momentum: $m$, number of layers in $S$: $l_1$, ${G}_1$: $l_2$, $G_2$: $l_3$, training epochs: $T$.}
	\KwResult{Optimized external and internal private feature extractors, and shared feature extractor.}
	
	\For{$j = 1$ \KwTo $N_s$}{
		Compute the number of batches for source domain $j$: $total\_batches_{s_j} \gets \left\lceil \frac{|\mathcal{D}_{s_j}|}{b} \right\rceil$\;
	}
	Compute the number of batches for the target domain: $total\_batches_t \gets \left\lceil \frac{|\mathcal{D}_t|}{b} \right\rceil$\;
	Compute the maximum number of batches across all domains: $total\_batches \gets \max_{j} \left(total\_batches_{s_j}, total\_batches_t \right)$\;
	
	\For{$epoch = 1$ \KwTo $T$}{
		\For{$batch = 1$ \KwTo $total\_batches$}{
			\For{$j = 1$ \KwTo $N_s$}{
				\If{$batch \leq total\_batches_{s_j}$}{
					Compute the $j$-th internal private feature from the source domain: $\mathcal{F}_{s_j}^{\text{in}} \gets P_j(S(\mathcal{D}_{s_j}))$\;
					Compute the $j$-th external private feature from the source domain: $\mathcal{F}_{s_j}^{\text{ex}} \gets G_j(\mathcal{D}_{s_j})$\;
				}
			}
			\If{$batch \leq total\_batches_t$}{
				Compute the $j$-th internal private feature from the target domain: $\mathcal{F}_{t_j}^{\text{in}} \gets P_j(S(\mathcal{D}_{t}))$\;
				Compute the $j$-th external private feature from the target domain: $\mathcal{F}_{t_j}^{\text{ex}} \gets G_j(\mathcal{D}_{t})$\;
			}
			
			Compute the LMMD loss for the $N_s$ external private feature extractors: $\mathcal{L}_{\text{lmmd}_1} \gets \frac{1}{N_s} \sum_{j=1}^{N_s}\hat{D}_{l}(\mathcal{F}_{s_j}^{\text{ex}}, \mathcal{F}_{t_j}^{\text{ex}})$\;
			Compute the fusion features from the source domain: $\mathcal{\bar{D}}_{s_j} \gets F\left(\mathcal{ {F}}_{s_j}^{ex}, \mathcal{ {F}}_{s_j}^{in}\right)$\; 
			Compute the fusion features from the target domain: $\mathcal{\bar{D}}_{t_j} \gets F\left(\mathcal{ {F}}_{t_j}^{ex}, \mathcal{ {F}}_{t_j}^{in}\right)$\;  
			Compute the LMMD loss for the fused features: $\mathcal{L}_{\text{lmmd}_2} \gets \frac{1}{N_s} \sum_{j=1}^{N_s}\hat{D}_{l}(\mathcal{\bar{D}}_{s_j}, \mathcal{\bar{D}}_{t_j})$\;
			Feed the fused features $\mathcal{\bar{D}}_{s_j}$ to the $j$-th classifier and obtain $C_j(\mathcal{\bar{D}}_{s_j})$; compute the classification loss $\mathcal{L}_{\text{cross}}$ as in Eq. (\ref{f7})\;
			Compute the classifier difference loss $\mathcal{L}_{\text{diff}}$ as in Eq. (\ref{f9})\;
			Update the external private feature extractors $G_1$ and $G_2$, the shared feature extractor $\rm S$, and the internal private feature extractors $P_1$ and $P_2$ by minimizing the total loss as in Eq. (\ref{f10})\;
		}
	}
\end{algorithm}

Once $\mathcal{L}_{total}$ converges to its minimum value, the corresponding AMDS-PFFA model parameters are saved. Subsequently, the target domain data $\mathcal{D}_t$ is input into the model to evaluate its performance.\par
\subsection{The specific models and parameters employed in each submodule of the AMDS-PFFA Framework for Drift Compensation}
In this work, we developed and implemented the architecture of each unit within the proposed AMDS-PFFA model framework, specifically designed for experimental validation under the condition of having two labeled source domains available at the initial stage of sensor array signal collection. The validation was performed in two real-world scenarios: the UCI drift gas experimental dataset, which covers a 36-month measurement period, and drift data from a gas sensor array collected using our self-developed E-nose systems over a 30-month period. The detailed architecture and parameters of each unit in the AMDS-PFFA framework are illustrated in Fig. \ref{model}. A comprehensive description of the architecture and parameters for each internal unit is provided below.\par
\subsubsection{The first stage - feature extractors}
as illustrated in Fig. \ref{model}, the model parameters are shown in black for the UCI drift data and in blue for the drift experimental data collected by the self-developed E-nose system. The shared feature extractor $S$ is composed of $N$ cascaded multi-head self-attention modules, followed by a two-layer fully connected neural network. For the UCI drift signal data, the first layer has 128 neurons, and the second layer has 128 neurons, while for the self-developed E-nose drift signal data, the first layer has 40 neurons, and the second layer has 32 neurons. This module employs weight-sharing to map all domains into a shared feature space.\par
The internal private feature extractors, $P_1$ and $P_2$, cascaded with the shared feature extractor $S$, are composed of a 1D-convolutional neural network (1D-CNN). For the UCI drift data, the convolutional layer has a kernel length of 3, a stride of 1, 128 input channels, 64 output channels, and "same" padding, producing a feature map of 64@1$\times$1. For the E-nose system drift data, the convolutional layer is configured with a kernel length of 3, a stride of 1, 32 input channels, 16 output channels, "same" padding, and outputs a feature map of 16@1$\times$1.\par
The external private feature extractors, $G_1$ and $G_2$, are similarly structured, consisting of a multi-head self-attention transformer followed by a two-layer fully connected neural network. For the UCI drift signal data, the first layer has 128 neurons, and the second layer has 64 neurons, while for the  self-developed E-nose system drift signal data, the first layer has 40 neurons, and the second layer has 16 neurons.\par

\subsubsection{The second stage - attentional feature fusion module}   
\begin{figure}[htbp]\centering
	\includegraphics[width=6.0cm]{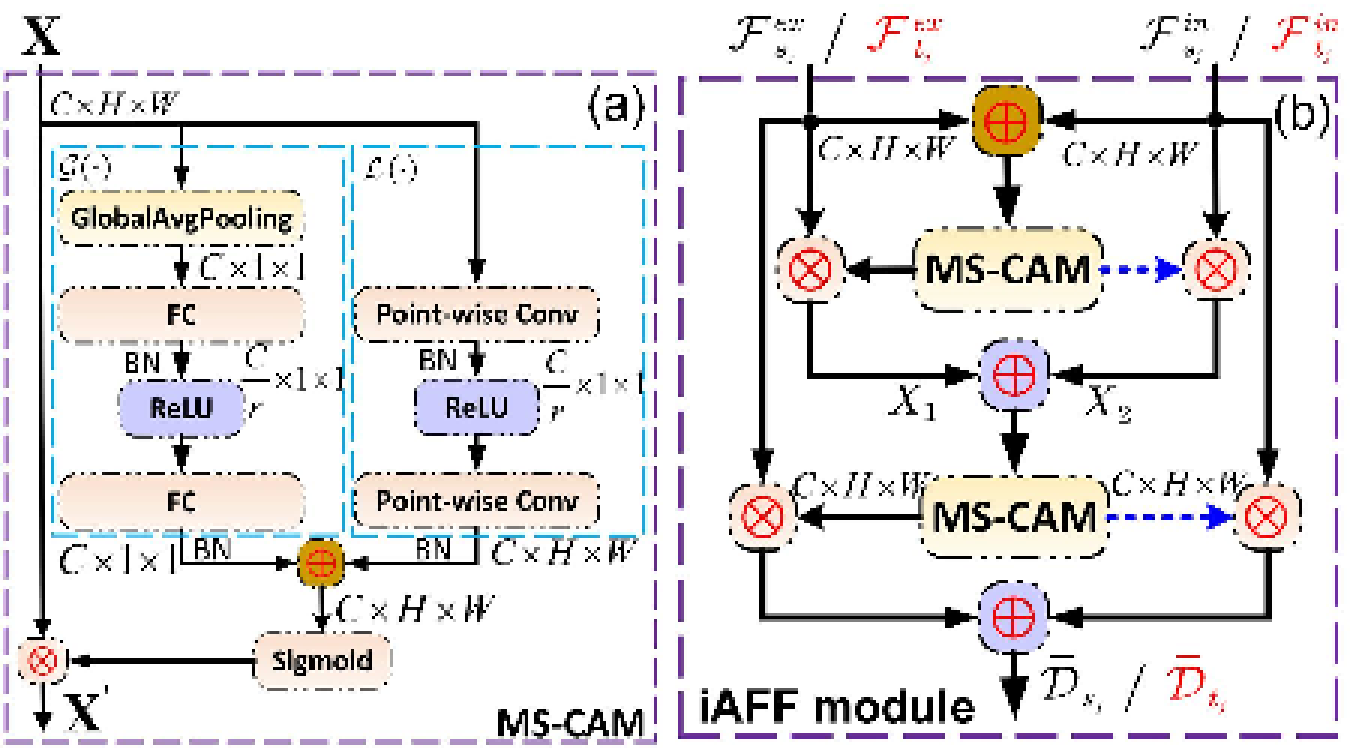}
	\caption{(a) The internal details of MS-CAM module and (b) the architecture of iAFF module.}
	\label {iAFF}
\end{figure}

In the second phase, we employ an attention-based feature fusion method designed for the characteristics of gas sensor array signals, referred to as iterative attentional feature fusion (iAFF) \cite{b33}. This method effectively merges the private feature outputs from both the external and internal private feature extractors. The architecture of the iAFF module and the detailed structure of the multi-scale channel attention module (MS-CAM) it employs are shown in Fig. \ref{iAFF} (a) and (b), respectively. The MS-CAM module includes a channel attention mechanism for global features, denoted as $\mathcal{G}(\cdot)$, and another for local features, denoted as $\mathcal{L}(\cdot)$. The output $X'$ of this module is computed as follows:\par
\begin{equation}\label{e10}
\small
\begin{aligned}
X'={X}\otimes \mathcal{Q} (X) = X\otimes sig(\mathcal{L}( {X})\oplus\mathcal{G}( {X}))
\end{aligned}
\end{equation}  
where $\mathcal{Q}(\cdot)$ denotes the input-output mapping function of MS-CAM, $sig(\cdot)$ refers to the sigmoid function, $\otimes$ denotes element-wise multiplication, and $\oplus$ denotes broadcasting addition. The computations for $\mathcal{L}(X)$ and $\mathcal{G}(X)$ are defined as:\par
\begin{equation}\label{e11}
\small
\begin{aligned}
\mathcal{L}(X) = \mathcal{B}\left(\mathcal{P}_2\left(\mathcal{B}\left(\mathcal{P}_1(X)\right)\right)\right)
\end{aligned}
\end{equation}
\begin{equation}\label{e12}
\small
\begin{aligned}
\mathcal{G}(X) = \mathcal{B}\left(\mathcal{P}_2\left(\mathcal{B}\left(\mathcal{P}_1\left(\text{avg}(X)\right)\right)\right)\right)
\end{aligned}
\end{equation}
where $avg(\cdot)$ denotes the global average pooling function, $\mathcal P_1$  and $\mathcal P_2$  denotes the first and second layers of point-wise convolution, respectively, and $\mathcal B$ denotes batch normalization operation. Thus, Formula (\ref{e10}) can be reformulated as:   \par 
\begin{equation}\label{e13}
\small
\begin{aligned}
X'& = X \otimes \mathcal{Q}(X) \\&= X \otimes \text{sig}\left(\mathcal{B}\left(\mathcal{P}_2\left(\mathcal{B}\left(\mathcal{P}_1(X)\right)\right)\right) \oplus \mathcal{B}\left(\mathcal{P}_2\left(\mathcal{B}\left(\mathcal{P}_1(\text{avg}(X))\right)\right)\right)\right)
\end{aligned}
\end{equation}
using the MS-CAM function $\mathcal{Q}(\cdot)$, the source domain fusion feature $\mathcal{\bar{D}}_{s_j}$, generated by the $j$-th external private feature $\mathcal{F}_{s_j}^{ex}$ and the $j$-th internal private feature $\mathcal{F}_{s_j}^{in}$ after passing through the iAFF module, is given by:\par
\begin{equation}\label{e14}
\small
\begin{aligned}
\mathcal{\bar{D}}_{s_j} = \mathcal{Q}\left(X_1 \oplus X_2\right) \otimes \mathcal{F}_{s_j}^{ex} + \left(1 - \mathcal{Q}\left(X_1 \oplus X_2\right)\right) \otimes \mathcal{F}_{s_j}^{in}
\small
\end{aligned}
\end{equation}
where The blue dashed arrowed line indicates $1-\mathcal{Q}(X_1\oplus X_2)$, and the initial integration $X_1 \oplus X_2$ in Eq. (\ref{e14}) is computed as:\par
\begin{equation}\label{e15}
\small
\begin{aligned}
X_1 \!\!\oplus\! X_2 \!\! =\! \!\mathcal{Q}\!\left(\mathcal{F}_{s_j}^{ex}\!+\!\mathcal{F}_{s_j}^{in}\right)\! \otimes\! \mathcal{F}_{s_j}^{ex} \!+\! \left(\!1 \!-\! \mathcal{Q}\!\left(\mathcal{F}_{s_j}^{ex}\!+\!\mathcal{F}_{s_j}^{in}\right)\!\right)\!\! \otimes\! \mathcal{F}_{s_j}^{in}
\end{aligned}
\small
\end{equation}\par
Similarly, the target domain fusion feature $\mathcal{\bar{D}}_{t_j}$, created by combining the $j$-th external private feature $\mathcal{F}_{t_j}^{ex}$ and the $j$-th internal private feature $\mathcal{F}_{t_j}^{in}$ after passing through the iAFF module, is given by:\par
\begin{equation}\label{e16}
\small
\begin{aligned}
\mathcal{\bar{D}}_{t_j} = \mathcal{Q}\left(X_1 \oplus X_2\right) \otimes \mathcal{F}_{t_j}^{ex} + \left(1 - \mathcal{Q}\left(X_1 \oplus X_2\right)\right) \otimes \mathcal{F}_{t_j}^{in}
\end{aligned}
\end{equation}
the expression for the initial integration $X_1 \oplus X_2$ in Eq. (\ref{e16}) is:\par
\begin{equation}\label{e17}
\small
\begin{aligned}
X_1 \! \!\oplus\! X_2 \!\! =\! \!\mathcal{Q}\!\left(\mathcal{F}_{t_j}^{ex}\!+\!\mathcal{F}_{t_j}^{in}\right)\! \otimes\! \mathcal{F}_{t_j}^{ex} \!+\! \left(\!1 \!-\! \mathcal{Q}\!\left(\mathcal{F}_{t_j}^{ex}\!+\!\mathcal{F}_{t_j}^{in}\right)\!\right)\!\! \otimes\! \mathcal{F}_{t_j}^{in}
\end{aligned}
\end{equation}
the input pairs for the iAFF module, namely $\mathcal{F}_{s_j}^{ex}$ and $\mathcal{F}_{s_j}^{in}$, as well as $\mathcal{F}_{t_j}^{ex}$ and $\mathcal{F}_{t_j}^{in}$, and their corresponding fused outputs $\mathcal{\bar{D}}_{s_j}$ and $\mathcal{\bar{D}}_{t_j}$, all have the same data size of $C \times H \times W$. For the UCI drift data, this size is $64 \times 1 \times 1$, while for our self-developed E-nose system, it is $16 \times 1 \times 1$. The parameter $r$ within the MS-CAM module is set to 1 in both cases.   
\subsubsection{The third stage - gas type classifiers} 
the gas type classifiers $C_1$ and $C_2$ each comprise a three-layer fully connected neural network. The first layer has neurons corresponding to the number of fused features obtained during the second stage, set at 64 for the UCI drift signal data and 16 for the drift signal from the self-developed E-nose system. The intermediate layer contains 25 neurons for the UCI drift data and 10 neurons for the self-developed E-nose drift data. The output layer has a number of neurons equal to the number of distinct gas types, which is 6 for the UCI drift signal data and 3 for the self-developed E-nose system drift data.\par	

\begin{figure*}[htbp]\centering
	\includegraphics[width=17.6cm]{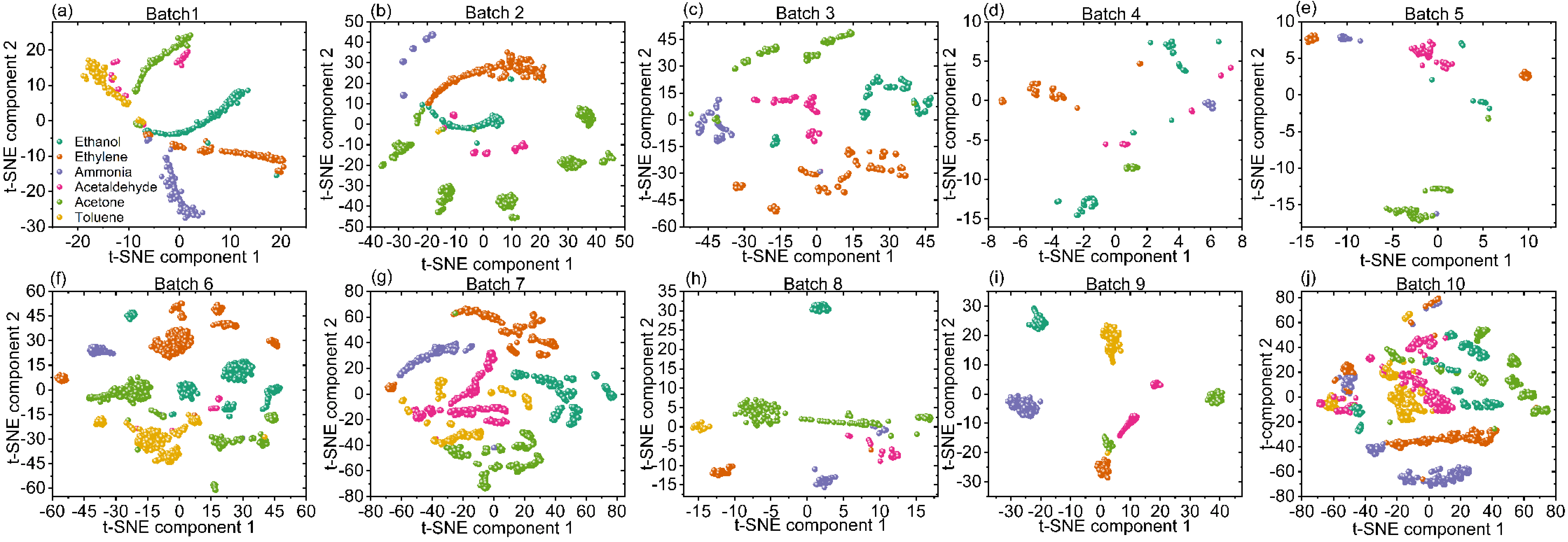}
	\caption{t-SNE-based 2D projection of 128-dimensional sensor array signals from batches 1 to 10. (a) Batch 1. (b) Batch 2. (c) Batch 3. (d) Batch 4. (e) Batch 5. (f) Batch 6. (g) Batch 7. (h) Batch 8.  (i) Batch 9. (j) Batch 10.}
	\label {tsne}
\end{figure*} 

\section{Experimental Validation of  Senosr Signal Drift Data from the UCI ChemoSignal Lab } 
\subsection{ The description of sensor signal drift data}
We first conducted a verification process using the gas sensor array drift dataset compiled by Alexander Vergara and Shankar Vembu from the ChemoSignals Laboratory at the BioCircuits Institute, University of California \cite{b34}. This dataset spans a 36-month period, from January 2007 to February 2011, and contains 13,910 measurements collected from 16 chemical sensor arrays. These sensors were employed to simulate drift compensation in a discrimination task involving six distinct gases, namely ethanol, ethylene, ammonia, acetaldehyde, acetone, and toluene, each present at varying concentrations. In each experiment, each sensor extracted 8 features that captured both the steady-state and dynamic response processes of the sensor signal. This resulted in a total of 128 features across all 16 sensors, denoted as $X=\{{x}_i\}_{i=1}^{|X|}$, where ${x}_i$ denotes an individual feature and $|X| = 128$ is the total number of features. The 36-month dataset is divided into 10 distinct batches, each corresponding to different time segments. A detailed overview of the gas sensor drift data is provided in Table \RNum{1}. Data labels are typically obtained during the initial stage of data collection. In this study, two labeled datasets from the initial stage, batches 1 and 2, are employed as source domains for experimental validation. These are referred to as source domain $n$ (where $n=1, 2$) in Fig. \ref{model}.
\begin{table}[htbp]
	\centering
	\caption{Details of the Gas Sensor Array Drift Dataset}
	\setlength{\tabcolsep}{0.5pt} 
	\renewcommand{\arraystretch}{1.6} 
	\fontsize{14}{15}\selectfont	
	\scalebox{0.38}{ 
		\begin{tabular}{cccccccccccc}
			\toprule
			\toprule
			\multirow{2}{*}{\textbf{Batch No.}} & \textbf{\begin{tabular}[c]{@{}c@{}}Concentration \\ range / ppmv\end{tabular}} & \textbf{Batch 1} & \textbf{Batch 2} & \textbf{Batch 3} & \textbf{Batch 4} & \textbf{Batch 5} & \textbf{Batch 6} & \textbf{Batch 7} & \textbf{Batch 8} & \textbf{Batch 9} & \textbf{Batch 10} \\ 
			\cmidrule(lr){2-12} 
			& \textbf{Month Period} & 1$\sim$2 & 3$\sim$10 & 11$\sim$13 & 14$\sim$15 & 16 & 17$\sim$20 & 21 & 22$\sim$23 & 24, 30 & 36 \\ 
			\midrule
			\textbf{Acetone} & 50$\sim$1000 & 90 & 164 & 365 & 64 & 28 & 514 & 649 & 30 & 61 & 600 \\ 
			\textbf{Acetaldehyde} & 5$\sim$500 & 98 & 334 & 490 & 43 & 40 & 574 & 662 & 30 & 55 & 600 \\ 
			\textbf{Ethanol} & 12$\sim$1000 & 83 & 100 & 216 & 12 & 20 & 110 & 360 & 40 & 100 & 600 \\ 
			\textbf{Ethylene} & 10$\sim$300 & 30 & 109 & 240 & 30 & 46 & 29 & 744 & 33 & 75 & 600 \\ 
			\textbf{Ammonia} & 10$\sim$600 & 70 & 532 & 275 & 12 & 63 & 606 & 630 & 143 & 78 & 600 \\ 
			\textbf{Toluene} & 10$\sim$100 & 74 & 5 & 0 & 0 & 0 & 467 & 568 & 18 & 101 & 600 \\ 
			\textbf{Total} & & 445 & 1244 & 1586 & 161 & 197 & 2300 & 3613 & 294 & 470 & 3600 \\ 
			\bottomrule
			\bottomrule
		\end{tabular}
	}
\end{table}
\begin{figure*}[htbp]
	\centering
	\subfigure[]{
		\begin{minipage}[t]{0.25\linewidth}
			\centering
			\includegraphics[width=4.4cm]{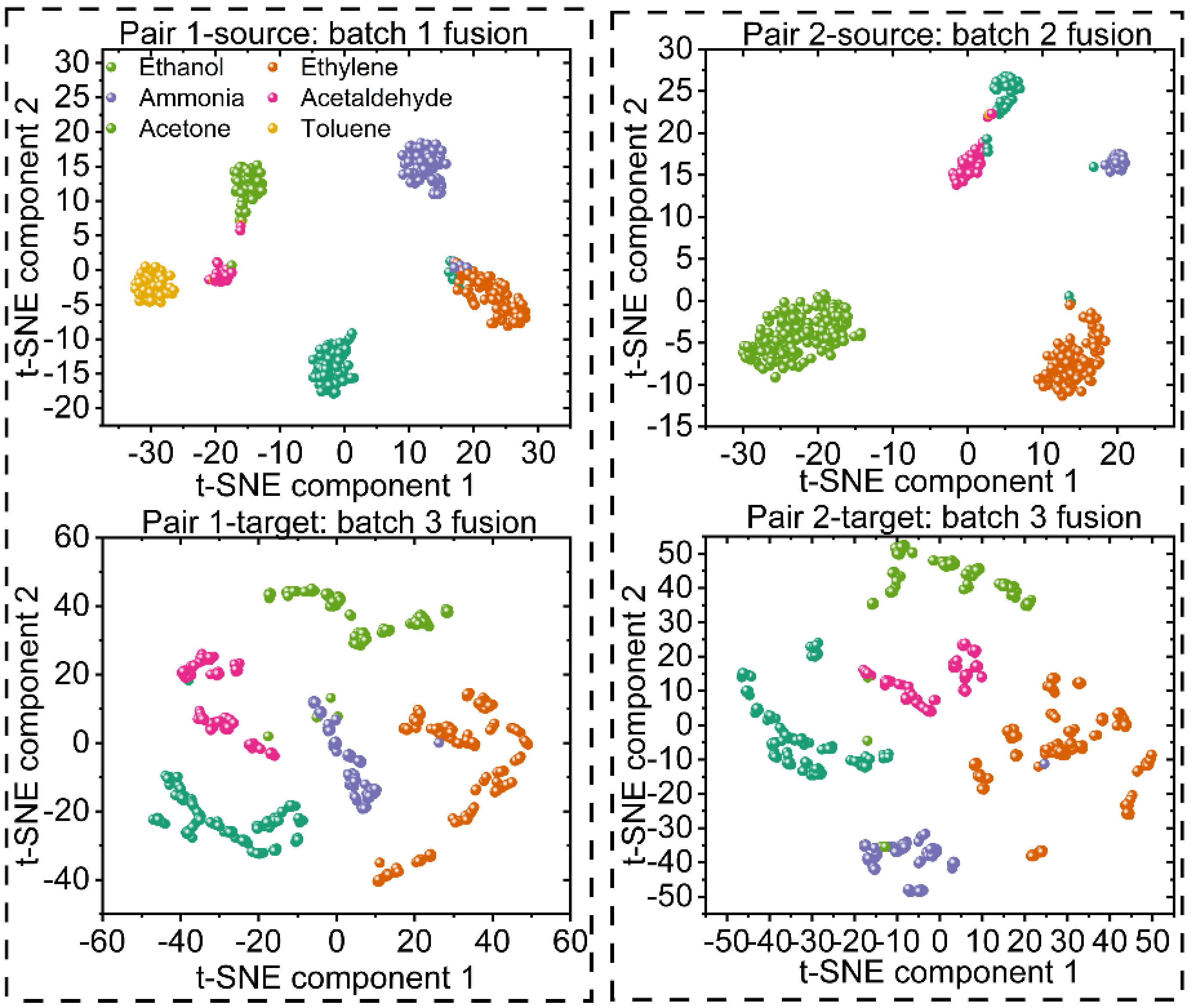}
		\end{minipage}%
	}%
	\subfigure[]{
		\begin{minipage}[t]{0.25\linewidth}
			\centering
			\includegraphics[width=4.4cm]{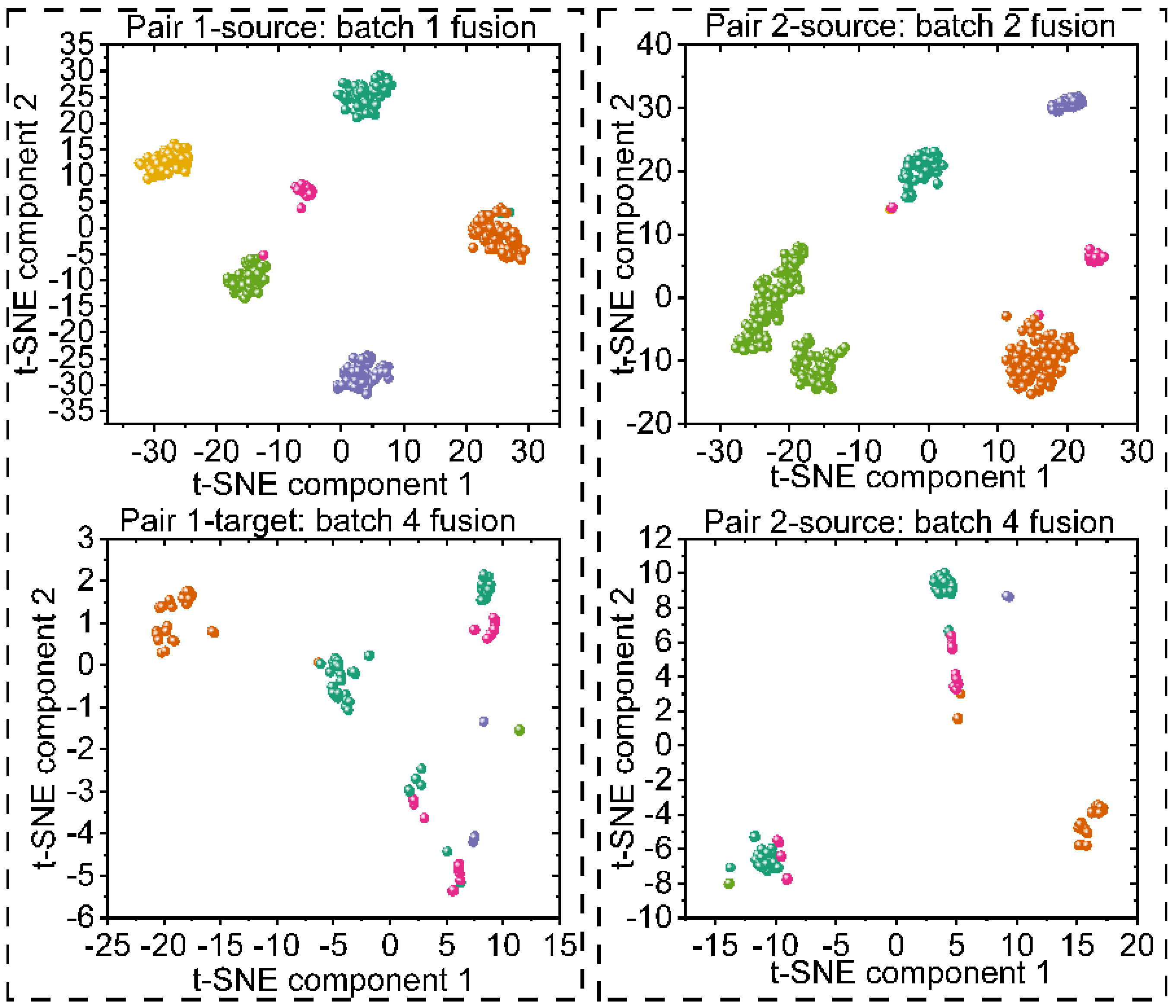}
		\end{minipage}%
	}%
	\subfigure[]{
		\begin{minipage}[t]{0.25\linewidth}
			\centering
			\includegraphics[width=4.4cm]{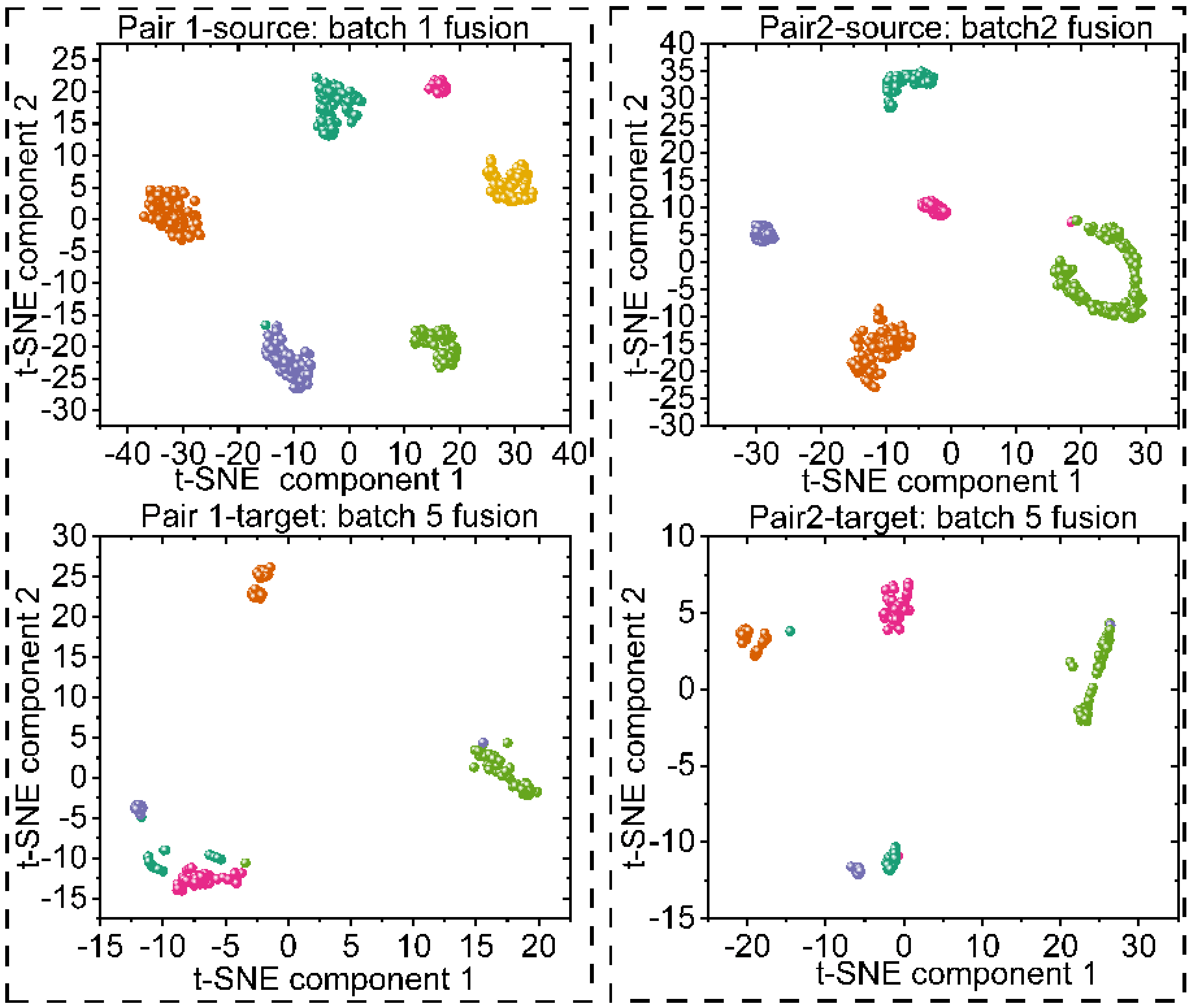}
		\end{minipage}%
	}%
	\subfigure[]{
		\begin{minipage}[t]{0.25\linewidth}
			\centering
			\includegraphics[width=4.4cm, height=3.7cm]{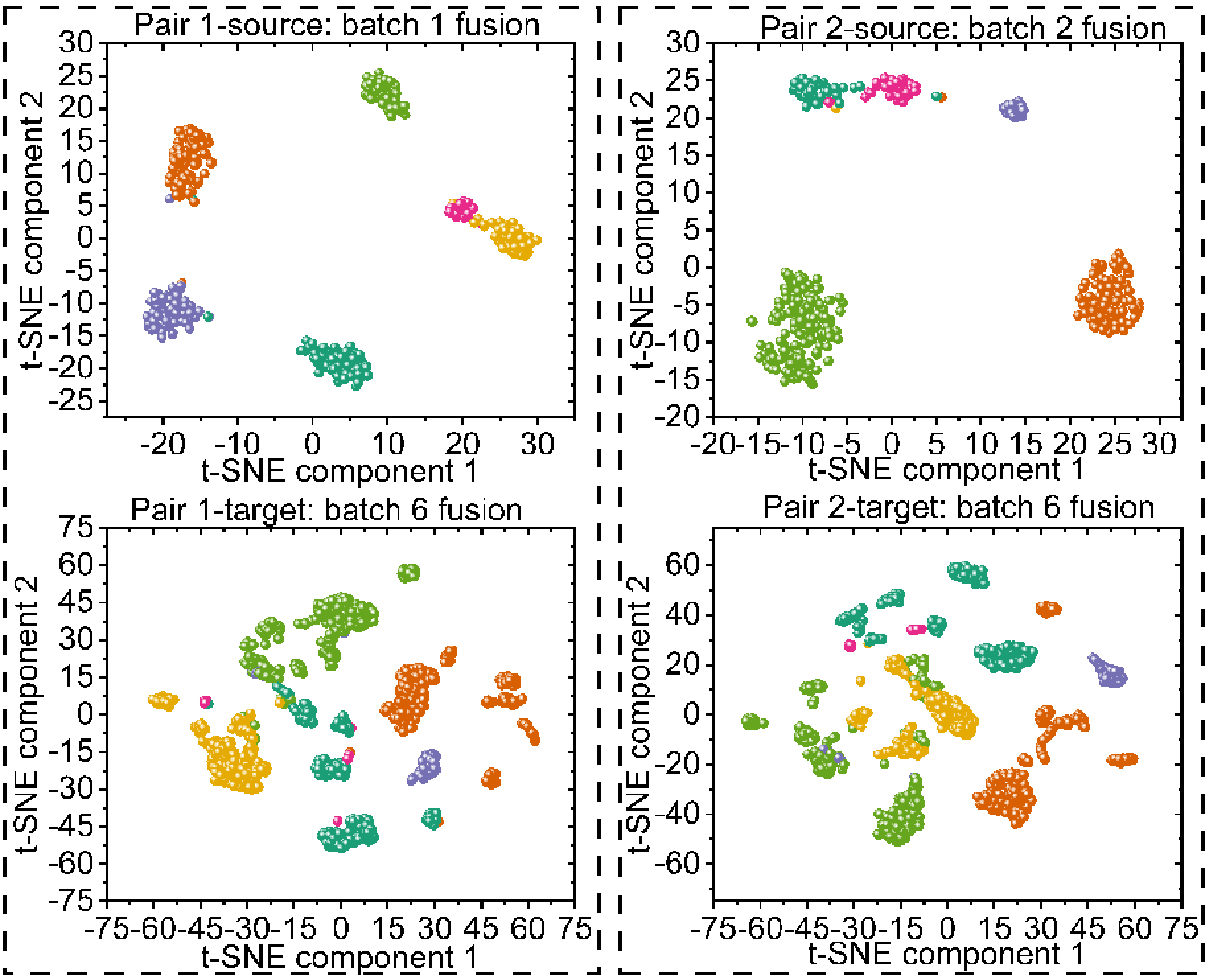}
		\end{minipage}%
	}%
	
	\subfigure[]{
		\begin{minipage}[t]{0.25\linewidth}
			\centering
			\includegraphics[width=4.4cm]{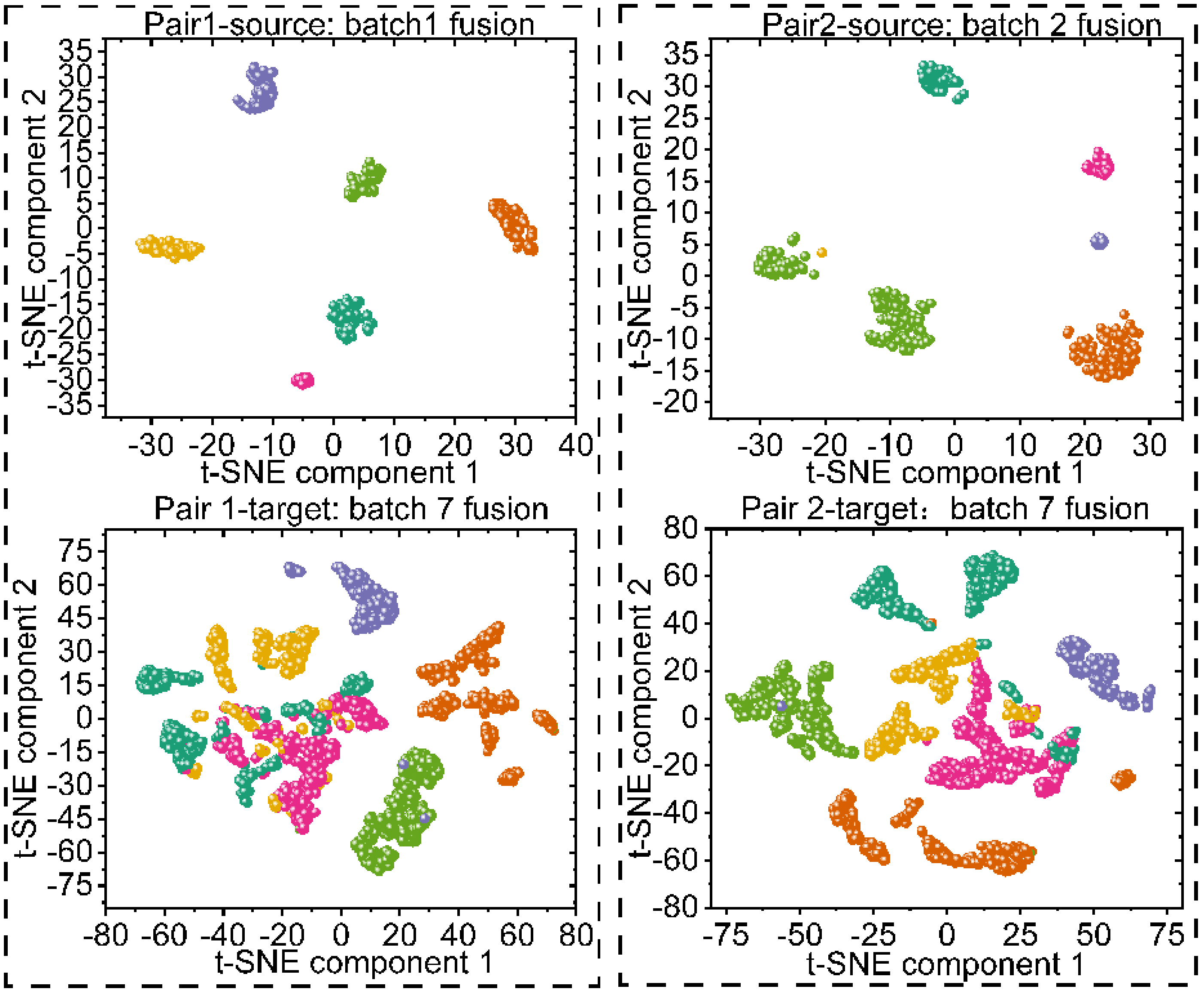}
		\end{minipage}%
	}%
	\subfigure[]{
		\begin{minipage}[t]{0.25\linewidth}
			\centering
			\includegraphics[width=4.4cm]{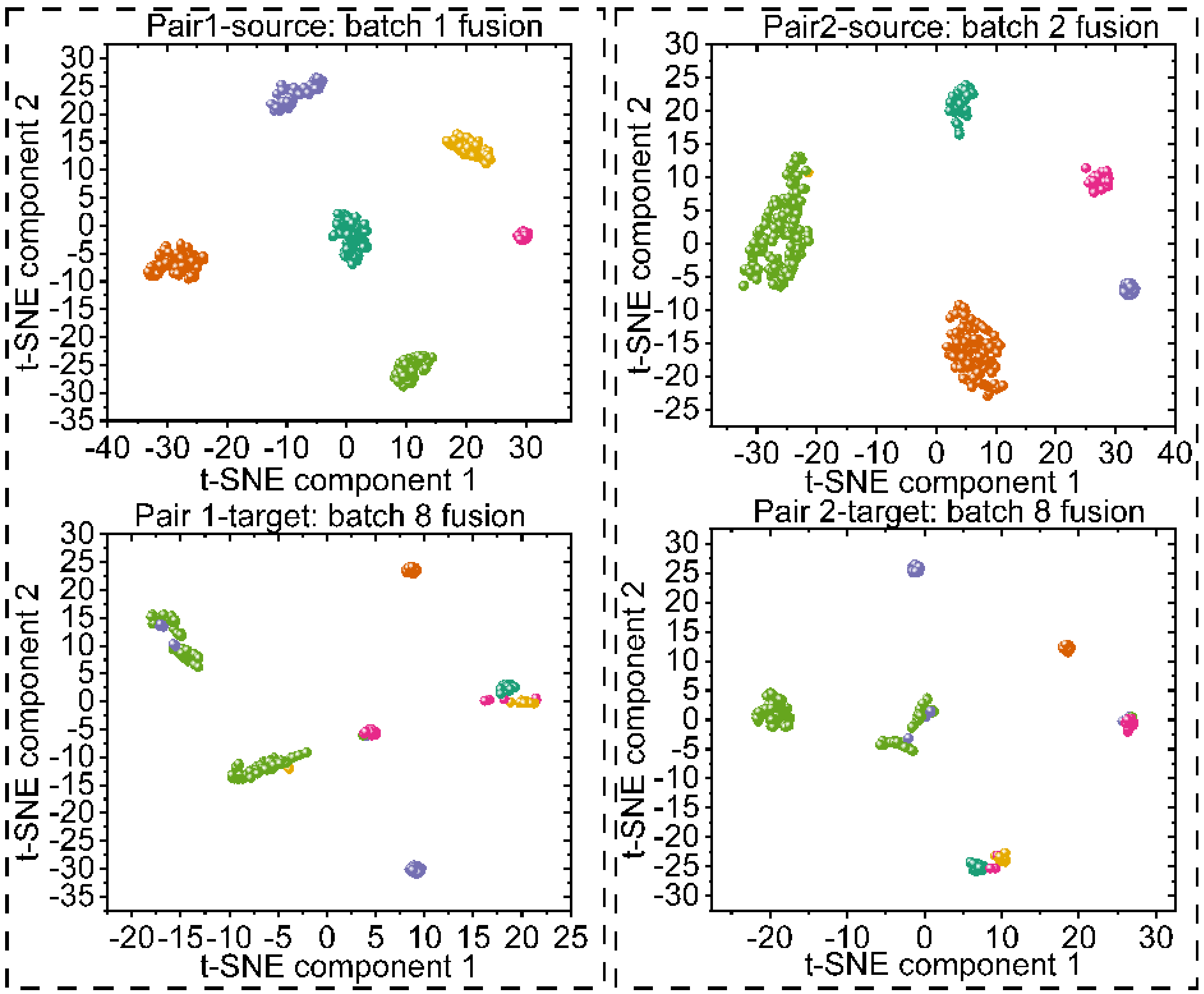}
		\end{minipage}%
	}%
	\subfigure[]{
		\begin{minipage}[t]{0.25\linewidth}
			\centering
			\includegraphics[width=4.4cm]{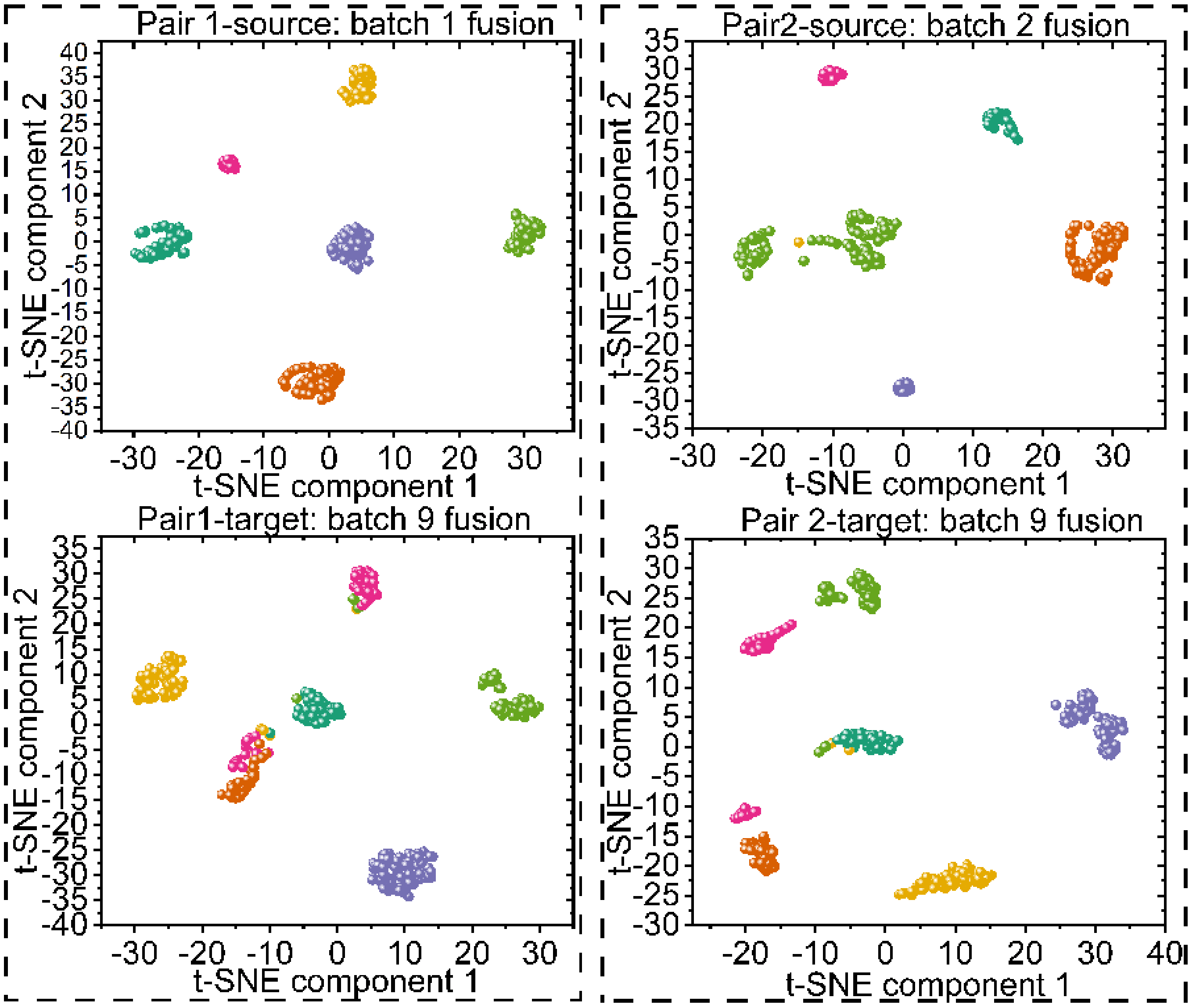}
		\end{minipage}%
	}%
	\subfigure[]{
		\begin{minipage}[t]{0.25\linewidth}
			\centering
			\includegraphics[width=4.4cm]{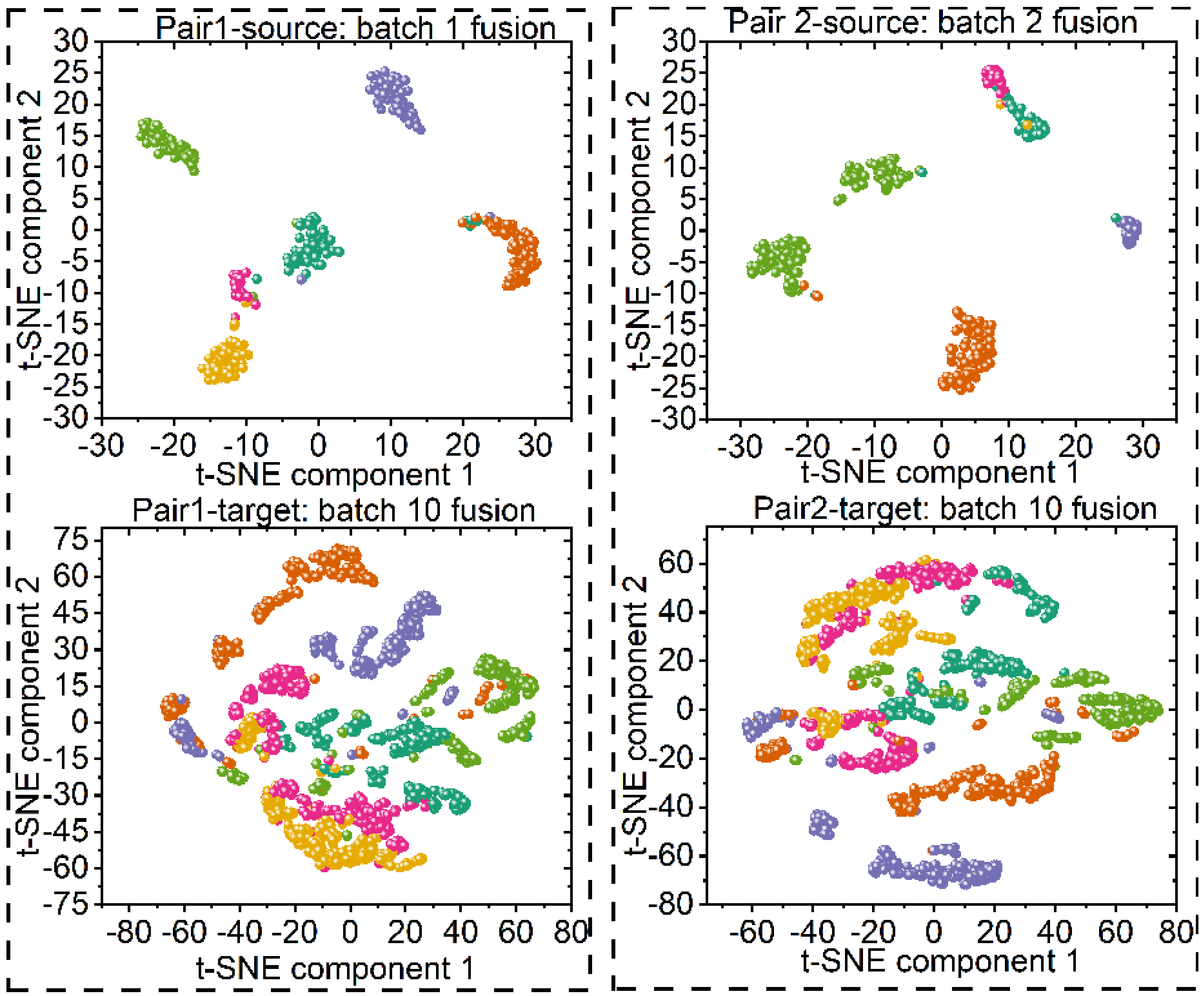}
		\end{minipage}%
	}%
	\caption{ Comparison of t-SNE 2D projections of fused features for source-target domain pairs after the iAFF module. (a-h) Source: batches 1 and 2; targets: batches 3 to 10.   }  
	\label {tsne1}
\end{figure*}  
\subsection{ The exprimental settings and results} 
The experimental setup for validating the AMDS-PFFA model is as follows: batch 1 and batch 2 serve as source domains, while each batch $k$ 
(where $k=3, 4, \dots, 10$) is employed sequentially as the target domain. Performance is evaluated on each target domain, and the results are recorded accordingly. \par
To observe the variations in data distribution across 10 different batches, we utilized the t-distributed stochastic neighbor embedding (t-SNE) method. This technique preserves the local structure of high-dimensional data, ensuring that similar data points remain close after dimensionality reduction. The resulting 2D distributions, derived from reducing the original 128-dimensional data, are shown in Fig. \ref{tsne} (a)-(j). From a global perspective, significant differences can be observed between the data distributions of source domains 1 and 2 and those of the target domains, with these differences becoming more pronounced over time. Locally, the distribution of different gas types varies notably across batches, with the data becoming increasingly dispersed as time progresses. This drifting behavior is both irregular and unpredictable, causing a substantial decline in the accuracy of the gas identifier trained on the initial labeled batch when applied to later drifted batches, often rendering it ineffective. \par
\begin{table}[htbp]
	\centering
\caption{Key Parameters of the AMDS-PFFA Model employed in the Experimental Setup.}
	\scalebox{0.64}{
		\begin{tabular}{ccccccccc}
			\toprule
			\toprule
			\multirow{-3}{*}{\textbf{Parameters}} & \shortstack{\textbf{1, 2} \\ \textbf{$\downarrow$} \\ \textbf{3}} & \shortstack{\textbf{1, 2} \\ \textbf{$\downarrow$} \\ \textbf{4}} & \shortstack{\textbf{1, 2} \\ \textbf{$\downarrow$} \\ \textbf{5}} & \shortstack{\textbf{1, 2} \\ \textbf{$\downarrow$} \\ \textbf{6}} & \shortstack{\textbf{1, 2} \\ \textbf{$\downarrow$} \\ \textbf{7}} & \shortstack{\textbf{1, 2} \\ \textbf{$\downarrow$} \\ \textbf{8}} & \shortstack{\textbf{1, 2} \\ \textbf{$\downarrow$} \\ \textbf{9}} & \shortstack{\textbf{1, 2} \\ \textbf{$\downarrow$} \\ \textbf{10}} \\
			\midrule
			\textbf{Coefficient $\alpha$} & 0.1 & 0.1 & 0.0008 & 0.5 & 0.1 & 2 & 0.01 & 0.5 \\
			\textbf{Learning rate} & 0.001 & 0.00075 & 0.002 & 0.003 & 0.009 & 0.006 & 0.009 & 0.001 \\
			\textbf{Batch size} & 48 & 12 & 8 & 48 & 48 & 36 & 48 & 36 \\
			\textbf{Weight decay} & $10^{-3}$ & $10^{-2}$ & $10^{-3}$ & $5\times10^{-4}$ & $10^{-3}$ & $10^{-3}$ & $10^{-2}$ & $5\times10^{-3}$ \\
			\textbf{Momentum} & 0.95 & 0.90 & 0.90 & 0.90 & 0.95 & 0.95 & 0.98 & 0.90 \\
			\textbf{Layers of S, G\textsubscript{1}, G\textsubscript{2}} & 3, 3, 3 & 3, 3, 3 & 1, 1, 1 & 3, 3, 3 & 3, 3, 3 & 2, 2, 2 & 1, 2, 2 & 2, 2, 2 \\
			\textbf{Dropout for S, G\textsubscript{1}, G\textsubscript{2}} & 0.3 & 0.2 & 0.01 & 0.3 & 0.01 & 0.15 & 0.01 & 0.1 \\	
			\bottomrule
			\bottomrule
		\end{tabular}
	}
	\label{para1}
\end{table}
\begin{table*}[htbp]
	\centering
	\caption{Comparison of gas identification accuracy (\%) for drift compensation models in the E-nose system.}
	\scalebox{0.650}{
		\begin{tabular}{@{}ccccccccccc@{}}
			\toprule
			\toprule
			\textbf{Model}      & \textbf{Year} & \textbf{1, 2 ---\textgreater{}3} & \textbf{1, 2 ---\textgreater{}4} & \textbf{1, 2 ---\textgreater{}5} & \textbf{1, 2 ---\textgreater{}6} & \textbf{1, 2 ---\textgreater{}7} & \textbf{1, 2 ---\textgreater{}8} & \textbf{1, 2 ---\textgreater{}9} & \textbf{1, 2 ---\textgreater{}10} & \textbf{\begin{tabular}[c]{@{}c@{}}Average \\  accuracy\end{tabular}} \\ \midrule
			\textbf{DAELM}       & 2015          & \underline{96.53}                        & 82.61              & 81.47                     & 84.97                        & \underline{71.89}                        & 78.10                        & \textbf{87.02}                       & 57.42                         & \underline{74.31}   \\ 
			\textbf{DRCA}       & 2017          & 92.69                        & \underline{87.58}               & \underline{95.90}                        & 86.52                        & 60.25                        & 62.24                        & 72.34                        & 52.00                         & 68.42                                                                \\ 
			\textbf{LME-CDSL}   & 2021          & 90.60                        & 68.32                        & 93.77                        & 73.22                        & 50.49                        & 60.54                        & \underline{79.89}                        & 41.66                         & 59.68                                                                \\ 
			\textbf{LDSP}       & 2022          & 96.78                        & 61.49                        & 93.90                        & 87.17                        & 58.54                        & 64.97                        & 78.09                        & 57.08                         & 69.98                                                                \\ 
			\textbf{DANP}      & 2022          & 95.80                        & 71.02                        & 95.20                        & 91.02                        & 65.54                        & 78.82                        & 77.14                        & 54.24                         & 72.25                                                                \\ 
			\textbf{TDACNN}     & 2022          & 83.83                        & 77.64                        & 75.63                        & 74.36                        & 62.08                        & 75.10                        & 60.85                        & 50.88                         & 64.60                                                                \\ 
			\textbf{M\textsuperscript{2}FL-CCC}   & 2023          & 89.73                        & 83.23                        & 75.64                        & 72.97                        & 61.36                        & 76.87                        & 61.28                        & 50.21                         & 64.83                                                                \\ 
			\textbf{DAEL-C}      & 2023          & 91.36                        & 75.95                        & 78.26                        & 71.37                        & 75.93                        & 49.81               & 67.51                        & 58.93                         & 71.15  
			\\ 
			\textbf{DAEL-D}      & 2023          & 90.42                        & 78.76                        & 72.13                        & 71.93                        & 63.72                        & 62.98               & 63.15                        & 43.19                         & 62.98  
			\\ 
			\textbf{DCLSL}      & 2023          & 94.76                        & 81.98                        & 94.92                        & 77.00                        & 63.36                        & \underline{85.03}               & 68.08                        & \underline{63.05}                         & 71.45                                                                \\ 
			\textbf{CDCNN}      & 2024          & 76.83                        & 72.44                        & 78.91                        & \underline{93.18}                        & 61.82                        & 71.52               & 56.64                        & 56.63                         & 68.59                                                                \\ 
			\textbf{AMDS-PFFA} & 2024          & \textbf{99.05}               & \textbf{88.46}                         & \textbf{98.44}                 & \textbf{93.31}               & \textbf{86.17}               & \textbf{89.24}                        & 71.06                        & \textbf{66.83}                & \textbf{83.20}  \\                                                     \bottomrule\bottomrule
	\end{tabular}}
	\label{acc1}
\end{table*}
\begin{figure*}[htbp]\centering
	\includegraphics[width=17.6cm]{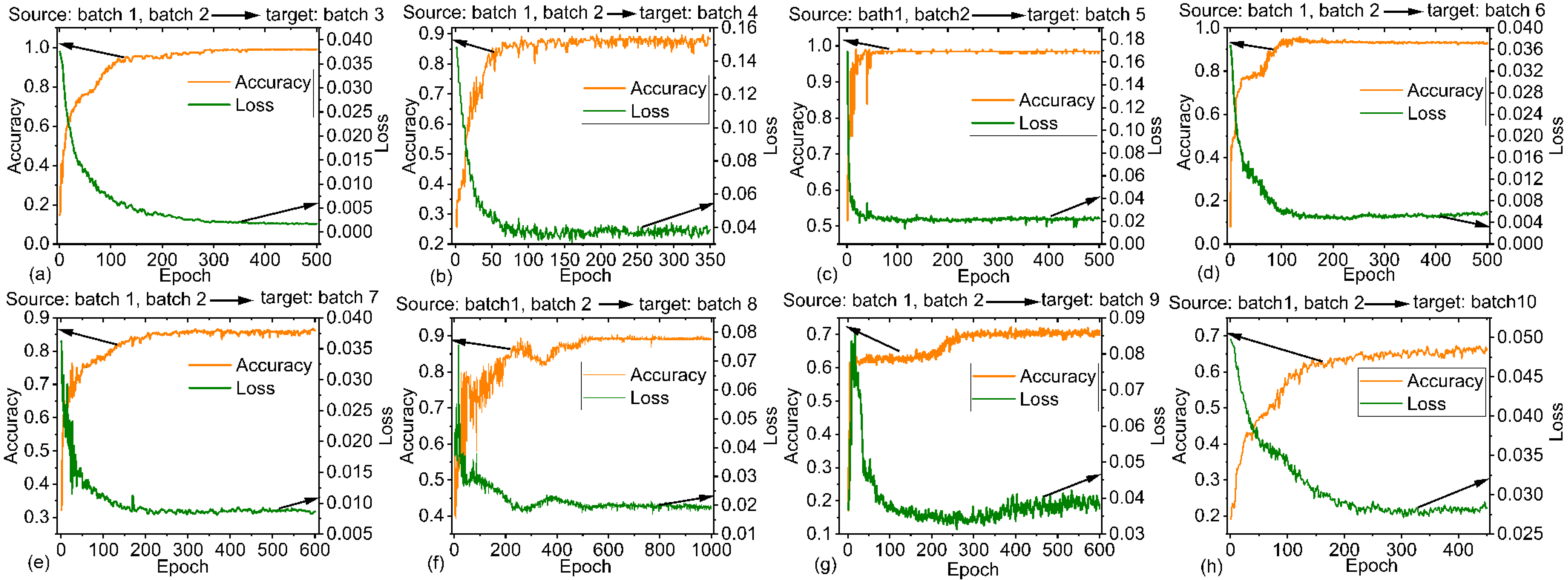}
	\caption{Convergence analysis of the unsupervised AMDS-PFFA model: accuracy and loss curves over epochs. (a) Target domain: batch 3. (b) Target domain: batch 4. (c) Target domain: batch 5. (d) Target domain: batch 6. (e) Target domain: batch 7. (f) Target domain: batch 8. (g) Target domain: batch 9. (h) Target domain: batch 10. }
	\label {acc_loss1}
\end{figure*}
The key parameters of the AMDS-PFFA model employed in the experimental setup are described in Table \ref{para1}. The term "layers of $S$, $G_1$, and $G_2$" refers to the number of multi-head self-attention modules within the shared feature extractor $S$ and the external private feature extractors $G_1$ and $G_2$. The experimental procedure is as follows: as shown in Fig. \ref{model}, batches 1 and 2 are utilized as source domains, while each batch $k$ (where $k = 3, 4, \ldots, 10$) is treated as the target domain. Algorithm 1 is then applied to optimize the model parameters until the total loss $\mathcal{L}_{total}$ reaches its minimum. Subsequently, the model parameters are saved, and target domain batch $k$ is employed for testing.\par 
To evaluate the efficacy of the unsupervised AMDS-PFFA model proposed in this work, which leverages labeled data from diverse source domains to mitigate gas sensor drift and enhance gas identification, we performed a comparative analysis with recently reported models introduced since 2015. The models compared include: 1) Cross-domain learning-based models: DAELM \cite{b36}, DRCA \cite{b25}, LDSP \cite{b26}, local manifold embedding cross-domain subspace learning (LME-CDSL) \cite{r25}, domain correction latent subspace learning (DCLSL)\cite{b27}, and DANP \cite{b29}; 2) deep learning-based models: target-domain-free domain adaptation convolutional neural network (TDACNN) \cite{r26}, multibranch multilayer feature learning and a comprehensive classification criterion (M\textsuperscript{2}FL-CCC)\cite{b37}, and contrastive domain generalization convolution neural network (CDCNN) \cite{c1}; and 3) active learning-based models: cross/discriminative domain-adaptation-based active ensemble Learning (DAEL-C, DAEL-D) \cite{b38}. Given that active learning models require a specific number of labeled target domains for effective guidance, we refer to the study \cite{b27} that employs 10 guidance samples for comparison.\par

The gas recognition accuracy of the AMDS-PFFA model, compared to these models across various target domain batches ($k = 3, 4, \ldots, 10$), is presented in Table \ref{acc1}. Due to varying sample sizes across batches, the average accuracy in our work is calculated based on the weighted distribution of sample numbers. The accuracy values shown in Table \ref{acc1} are averaged after reaching a steady state. The AMDS-PFFA model exhibited superior gas recognition performance compared to other models, achieving the highest accuracy in target domain batches 3, 4, 5, 6, 7, 8, and 10, with accuracies of 99.05\%, 88.46\%, 98.44\%, 93.31\%, 86.17\%, 89.24\%, and 66.83\%, respectively. The only exception was batch 9, where it did not achieve the highest accuracy. The model achieved an overall average gas identification accuracy of 83.17\%, the highest among those tested. Compared to the DCLSL method reported in 2023, the AMDS-PFFA model shows gas recognition accuracies that are 16.31\% and 22.81\% higher for target domain batches 6 and 7, respectively, and an overall average accuracy improvement of 11.75\%. This substantial enhancement in gas identification accuracy demonstrates the effectiveness of the unsupervised AMDS-PFFA model in mitigating gas sensor drift and improving gas identification precision. \par

The 2D distribution of fused features, reduced using the t-SNE method from the two source domains (batch 1 and batch 2) across each target batch, as processed through the iAFF fusion unit, is illustrated in Fig. \ref{tsne1}. Panels (a) through (h) correspond to the respective source-target pairs. Compared with Fig. \ref{tsne}, although some distribution differences exist between the source and target domain pairs, the overall data distribution is becoming increasingly consistent. Regarding the local distribution of various gas types, the alignment between the source and target domains is also improving, with the data becoming more concentrated and uniform.  This demonstrates that the proposed unsupervised AMDS-PFFA model effectively enhances the consistency of data distribution between the source and target domain pairs.\par 
\subsection{ Convergence study of the AMDS-PFFA model}
To demonstrate the convergence of the proposed AMDS-PFFA model within a limited number of training epochs, we recorded the accuracy and loss curves for each experiment, monitoring their changes as the number of epochs increased. The accuracy and loss curves for target domains 3 through 10 are presented in Fig. \ref{acc_loss1} (a) through (h), respectively. Generally, the accuracy and loss curves for each target domain, with the exception of batch 8, consistently converge and reach a relatively stable state within approximately 200 epochs. Although the curves for target domain batch 8 exhibit minor oscillations before epoch 400, they ultimately stabilize after 420 epochs of training. Overall, the experimental results confirm the robust convergence of the AMDS-PFFA model.\par  
\section{Experimental Validation of sensor signal Drift Data from a self-developed E-nose System } 
\subsection{ Self-developed E-nose system and drift signal data collection } 
\begin{figure*}[]
	\centering
	\subfigure[]{
		\begin{minipage}[t]{0.25\linewidth}
			\centering
			\includegraphics[width=4.4cm,height=3.1cm]{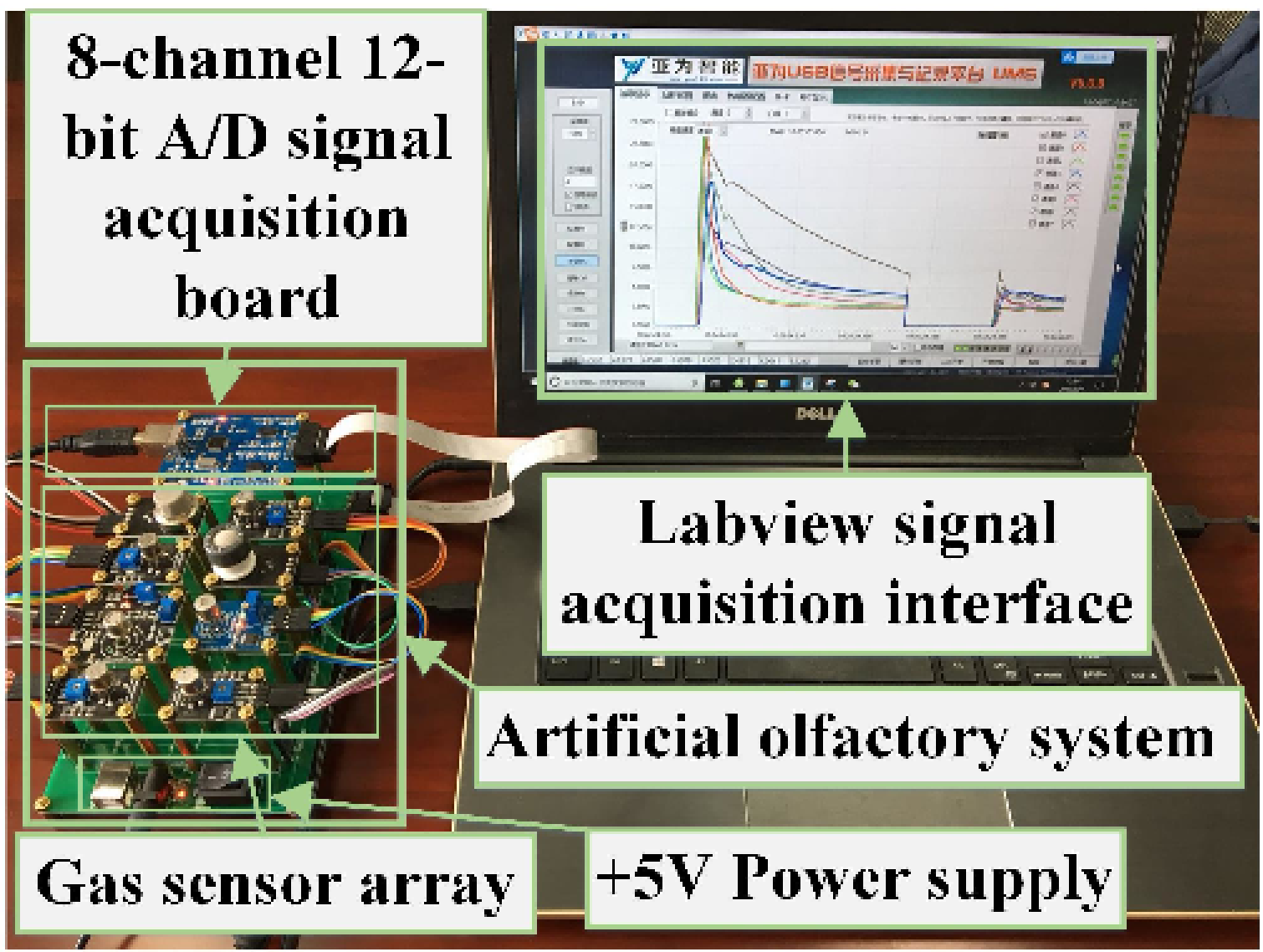}
		\end{minipage}%
	}%
	\subfigure[]{
		\begin{minipage}[t]{0.25\linewidth}
			\centering
			\includegraphics[width=2.3cm,height=3.1cm]{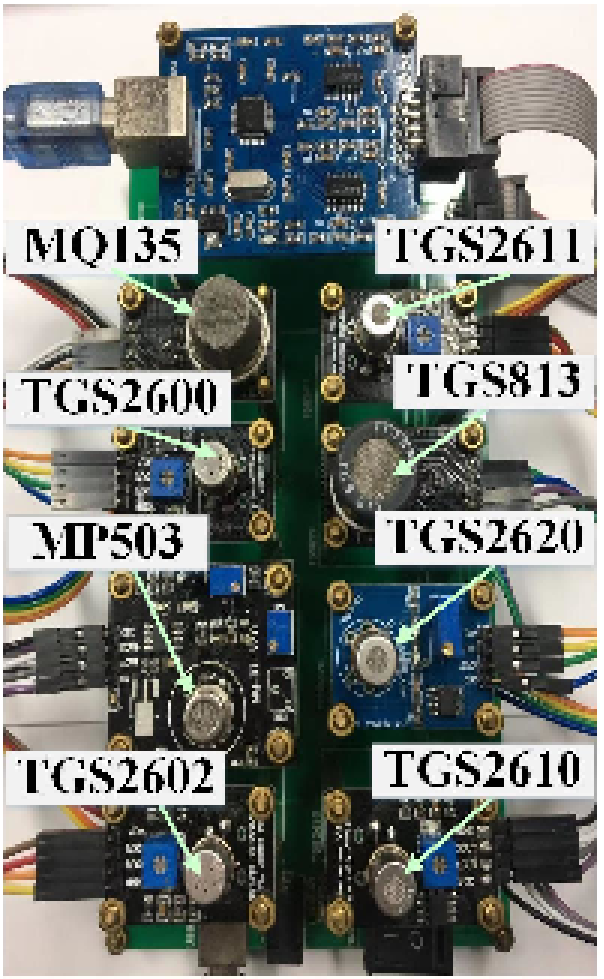}
		\end{minipage}%
	}%
	\subfigure[]{
		\begin{minipage}[t]{0.25\linewidth}
			\centering
			\includegraphics[width=4.4cm,height=3.1cm]{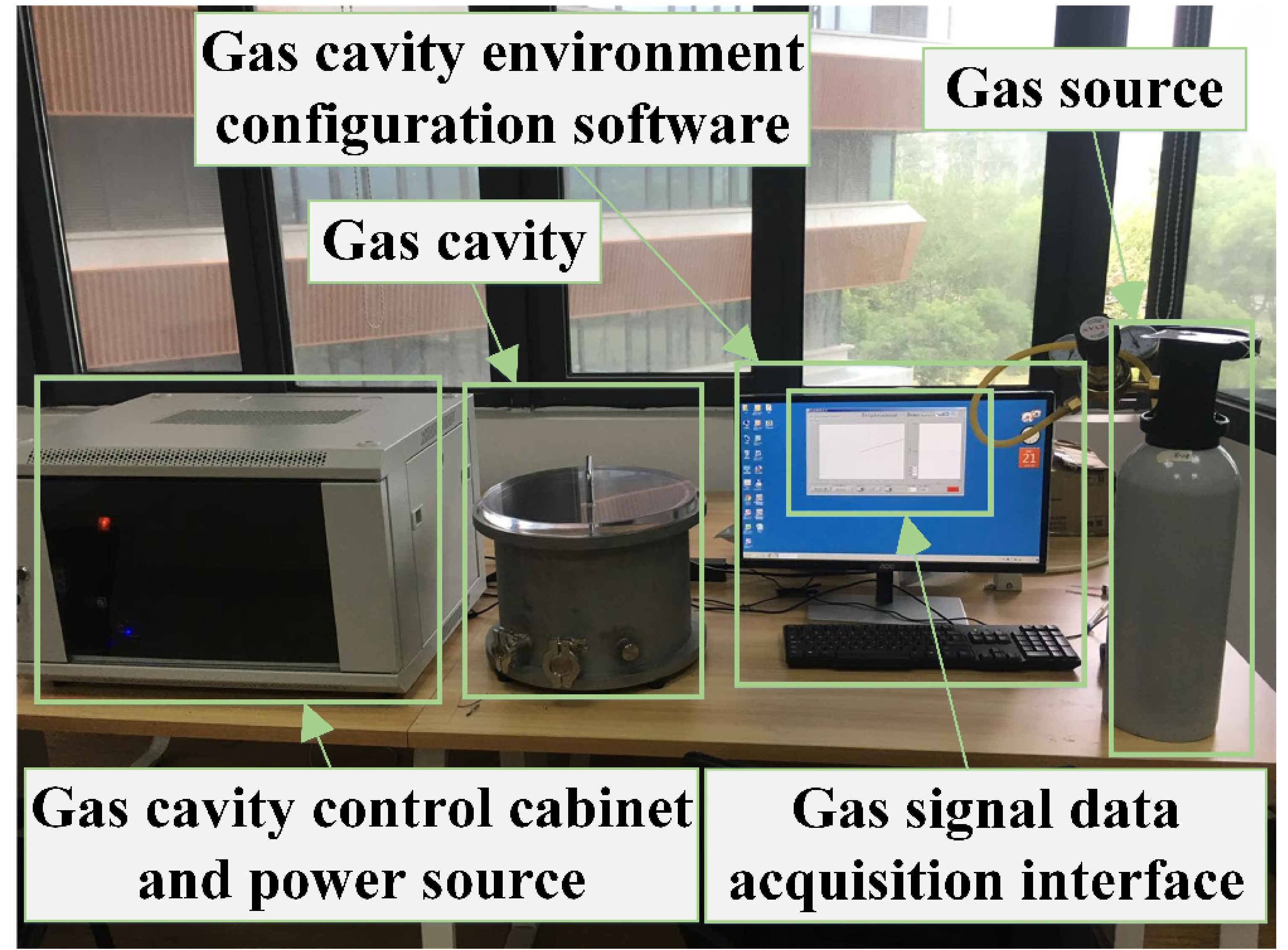}
		\end{minipage}%
	}%
	\subfigure[]{
		\begin{minipage}[t]{0.25\linewidth}
			\centering
			\includegraphics[width=3.0cm,height=3.1cm]{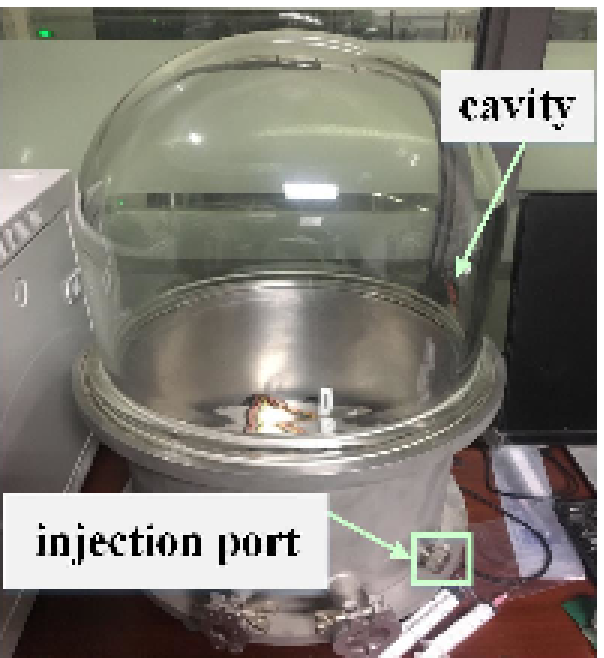}
		\end{minipage}%
	}%
	\centering
	\caption{(a) The self-developed E-nose system. (b) Gas sensor array. (c) The full suite of experimental gas distribution equipment. (d) 35 L gas distribution chamber.} 
	\label {Enose}
\end{figure*}

To further demonstrate the capability of the proposed unsupervised attention-based AMDS-PFFA model in mitigating drift and enhancing gas identification accuracy, we fabricated a custom E-nose system in our laboratory. The custom-designed  E-nose system and the gas sensor array with a 8-channel 12-bit A/D signal acquisition board  are  illustrated in Fig. \ref{Enose} (a) and (b), respectively. We selected toxic, harmful, flammable, and explosive gases, specifically CO,  H\textsubscript{2}, and mixed gases of CO and H\textsubscript{2}, for target identification. In accordance with these target gas, we selected 8 distinct gas sensors including Figaro's tgs813, tgs2611, tgs2610, tgs2620, tgs2600 and tgs2602  and Winsen's  mp503 and mq135 to compose the sensor array.\par 
The experiment employs the syringe static injection method. The entirety of the experimental gas distribution equipment and gas distribution chamber are illustrated in Fig. \ref{Enose} (c) and (d), respectively. The signal acquisition frequency is set to 10 Hz.   
Detailed experimental procedures are provided in the supplementary material.\par
\begin{figure*}[htbp]\centering
	\includegraphics[width=17.6cm]{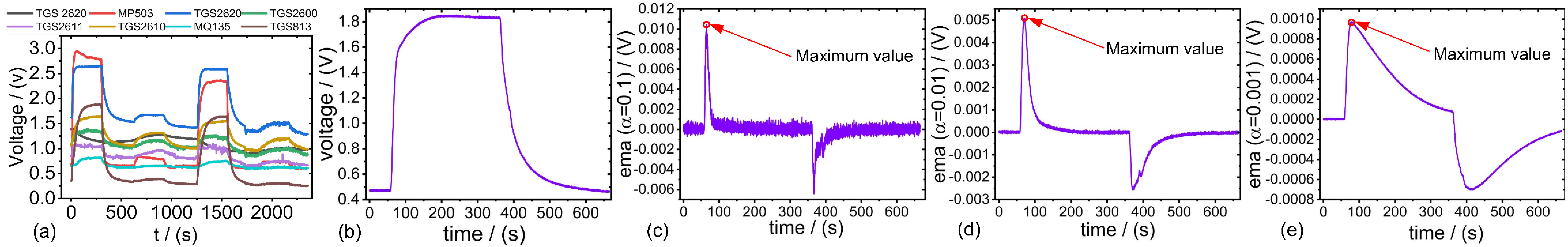}
	\caption{(a) Real-time response signals from the E-nose continueouly monitoring single 600 ppm H\textsubscript{2}, single 600 ppm CO, a gas mixture of 400 ppm H\textsubscript{2} and 300 ppm CO, and single 300 ppm CO. (b) Real-time response signal of the tgs2610 at the gas mixture with 180 ppm CO and 180 ppm  H\textsubscript{2}. (c) ema signal with a smoothing factor $\alpha$=0.1. (d) $\alpha$=0.01. (e) $\alpha$=0.001.}
	\label {response}
\end{figure*}

During each experiment, a cumulative total of approximately 3500 $\times$8 sensor array signal points can be acquired. The signal response curves depicting the continuous recording of single 600 ppm H\textsubscript{2}, single 600 ppm CO, a gas mixture of 400 ppm H\textsubscript{2} and 300 ppm CO, and single 300 ppm CO are illustrated in Fig. \ref{response} (a). Details regarding the drift data of gas sensor array signals obtained from our self-developed E-nose system are described in Table \ref{details_self}. The comprehensive experiment extended over a duration of 30 months.            \par

\begin{table}[h]
	\centering
	\caption{The details of gas sensor array drift data acquired from self-developed E-nose system.}
	\scalebox{0.60}{
		\begin{tabular}{cccccc}
			\toprule
			\toprule
			\multirow{3}{*}{\textbf{Gas types}} & \multirow{3}{*}{\begin{tabular}[c]{@{}c@{}}\textbf{Concentration} \\ \textbf{range/ ppm}\end{tabular}} & \multicolumn{4}{c}{\textbf{Experimental time}} \\
			\cmidrule{3-6}
			& & \multicolumn{1}{c}{\textbf{2020.06}} & \multicolumn{1}{c}{\textbf{2021.06}} & \multicolumn{1}{c}{\textbf{2022.03}} & \textbf{2022.12} \\
			\cmidrule{3-6}
			& & \multicolumn{1}{c}{\textbf{Batch 11}} & \multicolumn{1}{c}{\textbf{Batch 12}} & \multicolumn{1}{c}{\textbf{Batch 13}} & \textbf{Batch 14} \\
			\midrule
			\textbf{CO} & 0$\sim$1000 ppm & 118 & 199 & 149 & 163 \\
			\textbf{H\textsubscript{2}} & 0$\sim$1000 ppm & 115 & 199 & 149 & 172 \\
			\begin{tabular}[c]{@{}c@{}}\textbf{CO and} \\ \textbf{H\textsubscript{2}} \\ \textbf{gas mixture}\end{tabular} & \begin{tabular}[c]{@{}c@{}}CO: 0$\sim$500ppm \\ H\textsubscript{2}: 0$\sim$200ppm\end{tabular} & 120 & 240 & 167 & 173 \\
			\midrule
			\textbf{Total} & & 353 & 638 & 465 & 508 \\
			\bottomrule
			\bottomrule
		\end{tabular}
	}
	\label{details_self}
\end{table}
\begin{figure*}[htbp]
	\centering
	\subfigure[]{
		\begin{minipage}[t]{0.33\linewidth}
			\centering
			\includegraphics[width=5.5cm]{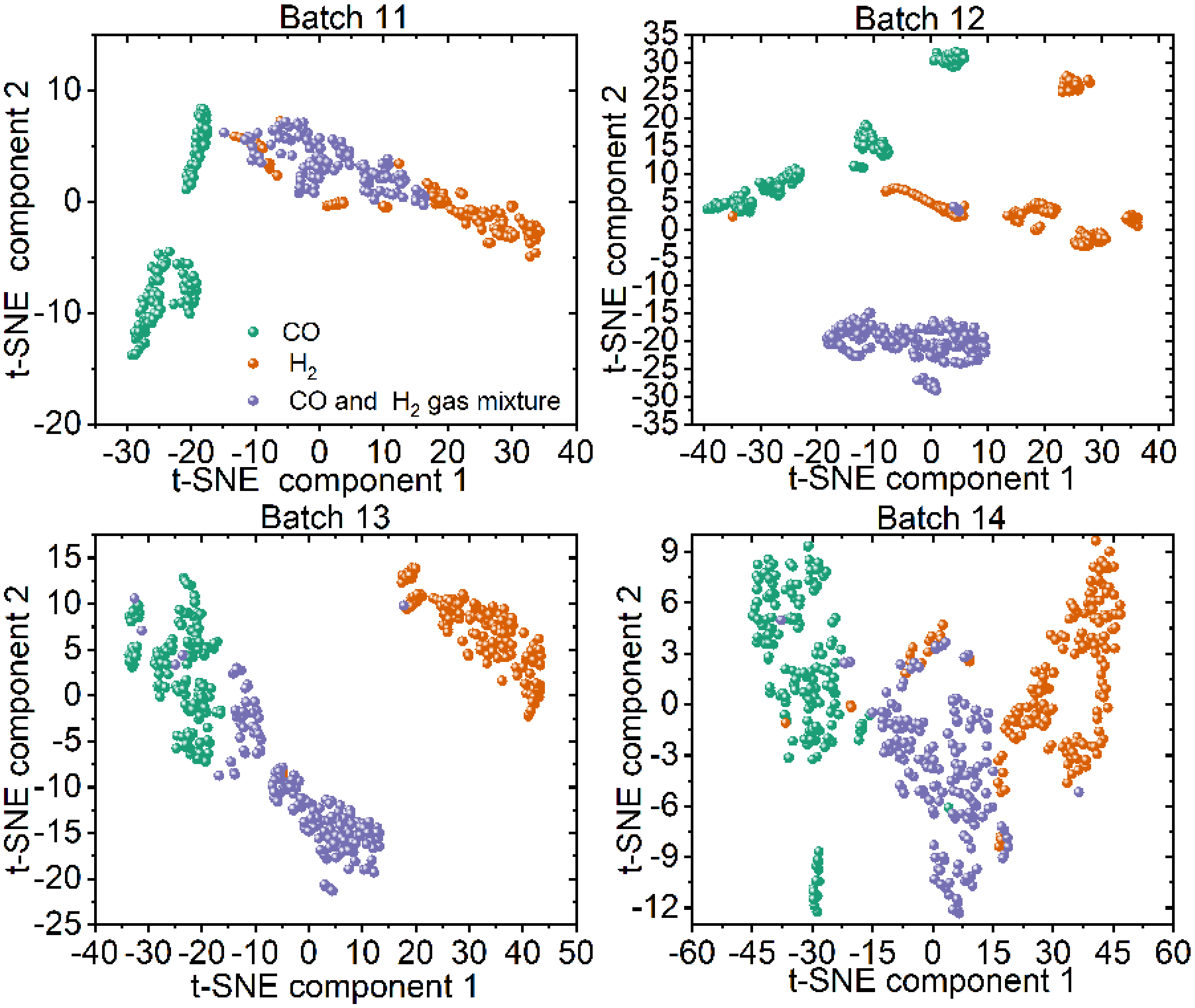}
		\end{minipage}%
	}%
	\subfigure[]{
		\begin{minipage}[t]{0.33\linewidth}
			\centering
			\includegraphics[width=5.5cm]{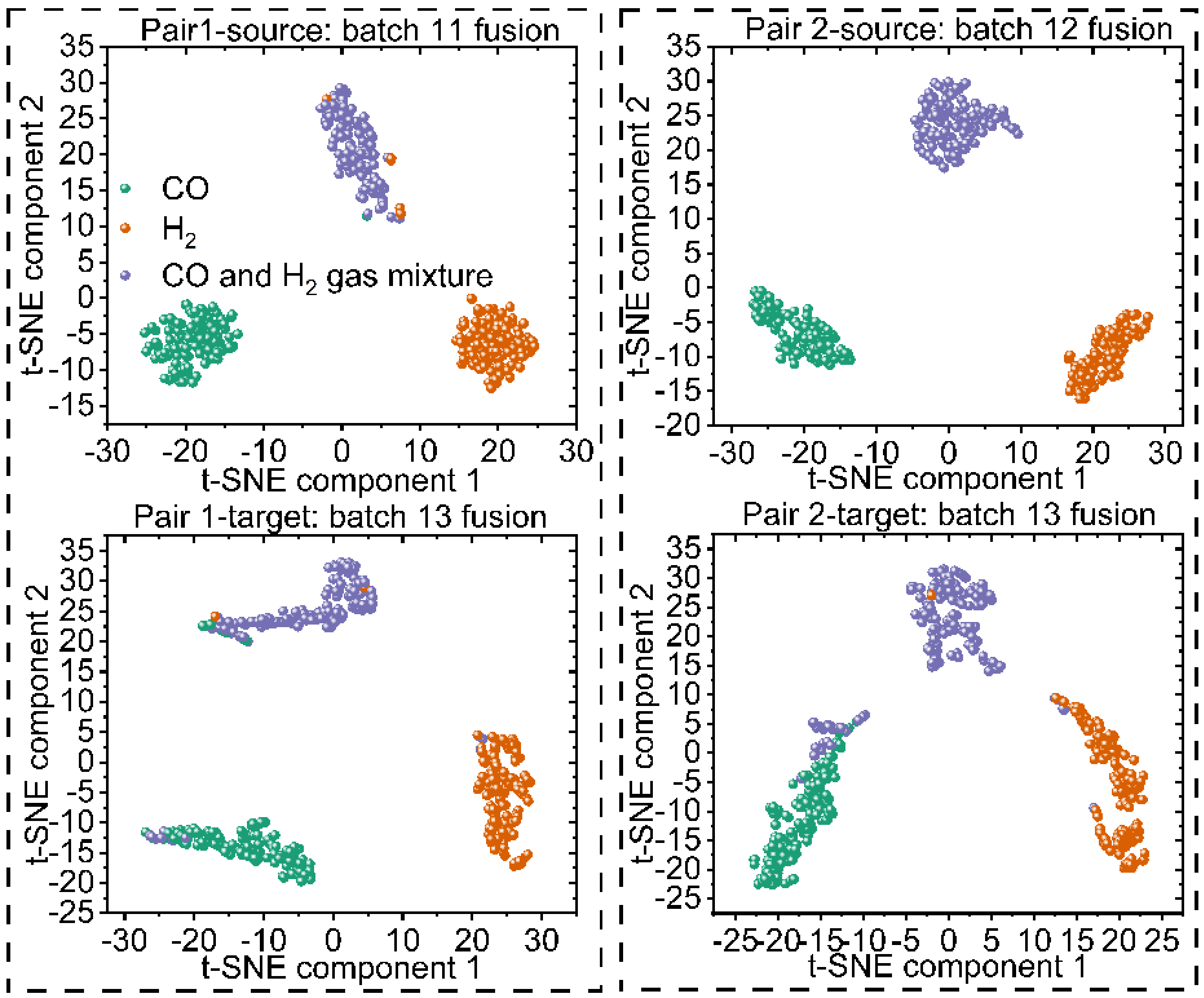}
		\end{minipage}%
	}%
	\subfigure[]{
		\begin{minipage}[t]{0.33\linewidth}
			\centering
			\includegraphics[width=5.5cm]{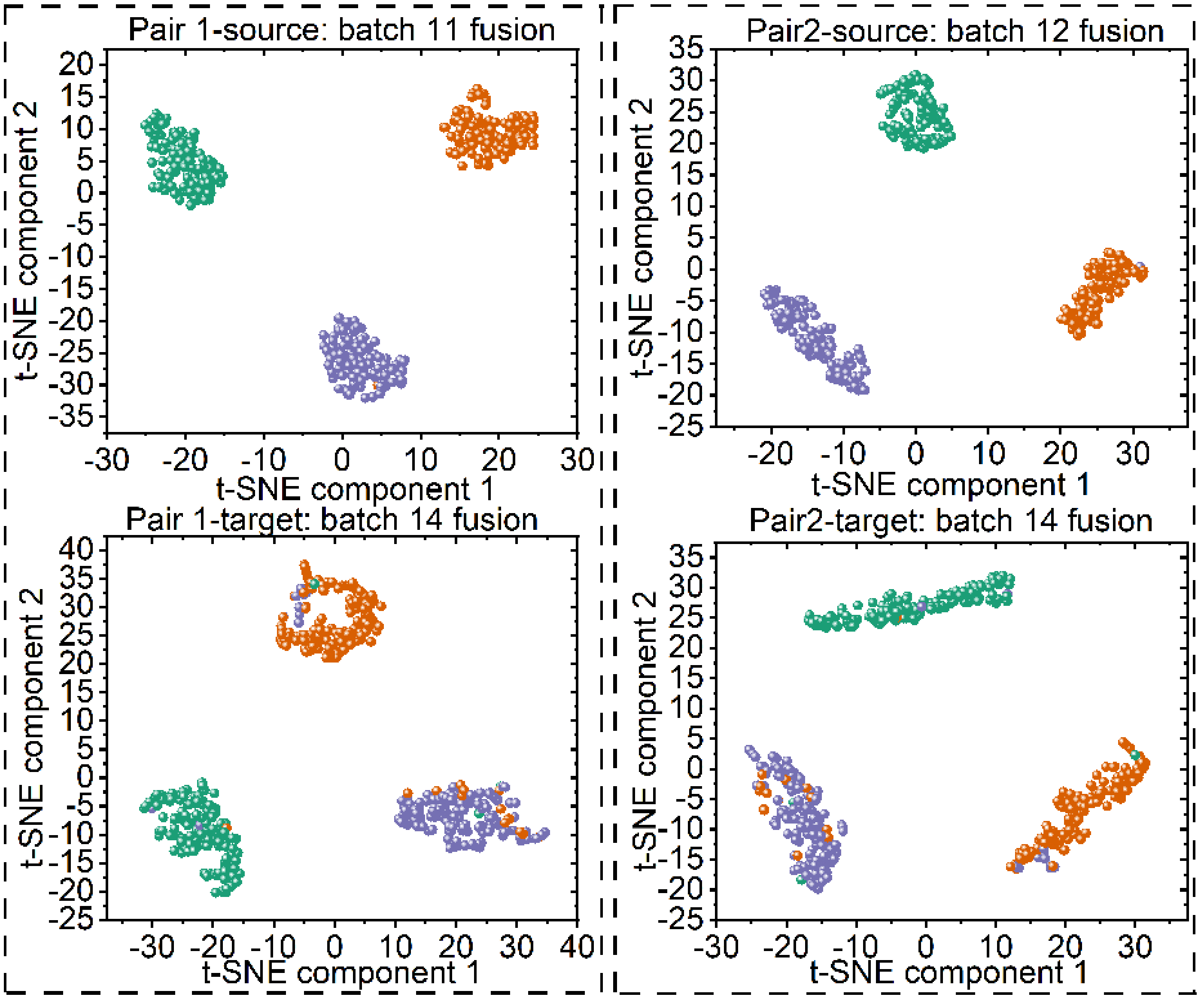}
		\end{minipage}%
	}%
	\caption{t-SNE 2D projections of sensor array signals and fused features after the iAFF module. (a) Projection of 40-dimensional signals from batches 11 to 14. (b-c) Comparison of source-target domain pairs: Source: batches 11-12, Target: batches 13-14. }  
	\label {fused_feature_compared2}
\end{figure*}
\subsection{Experimental validation of the model using drift data from the self-developed E-nose system}  
Extracting features from the dynamic response signal, consisting of $3500\times 8$ data points obtained from the sensor array, is a crucial step in gas recognition. The extracted features should preserve the significant effective identification information. Since the metal oxide semiconductor (MOS) gas sensor exhibits a sluggish reaction to gas,  we extract two steady-state features and three dynamic features.  The discrepancy between the mean steady-state value of the sensor's dynamic response signal and the baseline value is defined as the first steady-state feature value. Its  calculation expression is as follows:\par
\begin{equation}\label{f}
\small
\begin{aligned}
V_{diff}= \frac{1}{n-m+1}\sum_{k=m}^{n}V[k]-V_{baseline}
\end{aligned}
\end{equation}
where  $m$ and $n$ denote the commencement and conclusion points where the signal attains its steady-state, and $V[k]$ and $V_{baseline}$ denote the voltage at discrete point $k$ of the dynamic response signal and the recorded baseline voltage, respectively. The normalization of $ V_{diff}$ is selected as another steady-state feature, denoted as $V_{norm}$. Its expression is as follows:\par
\begin{equation}\label{f19}
\small
\begin{aligned}
V_{norm}= \frac{\frac{1}{n-m+1}\sum\limits_{k=m}^{n}V[k]-V_{baseline}}{V_{baseline}}
\end{aligned}
\end{equation}\par 
To more effectively capture fluctuations in the dynamic response trend of the signal. We compute the difference sequence of the dynamic response signal from the sensor, denoted as $\Delta V[k]$, where $\Delta V[k]=V[k]-V[k-1]$ , and subsequently determine the exponential moving average (EMA) sequence based on this difference sequence. The expression is as follows:\par
\begin{equation}\label{f}
\small
\begin{aligned}
ema_{\alpha}[k]=(1-\alpha)ema_{\alpha}[k-1]+\alpha(V[k]-V[k-1])
\end{aligned}
\end{equation}
where the initial value of the exponential moving average series, ema[0], is set to 0, and $\alpha$ denotes the smoothing coefficient. We set the values of $\alpha$ as 0.1, 0.01, and 0.001, respectively, and derive corresponding new sequence: $ema_{\alpha=0.1}[k]$, $ema_{\alpha=0.01}[k]$, $ema_{\alpha=0.001}[k]$, the maximum value within the resulting difference sequences $max(ema_{\alpha=0.1}[k])$, $max(ema_{\alpha=0.01}[k])$, $max(ema_{\alpha=0.001}[k])$ are employed as the dynamic-state features. We utilize the real-time response signal of the tgs2620 at the gas mixture with 180 ppm CO and 180 ppm $\rm H_2$ as an illustration,  its signal is depicted in Fig. \ref{response} (b), when the smoothing coefficient $\alpha$ is set to 0.1, 0.01 and 0.001, the resulting EMA signal is depicted in Fig. \ref{response} (c), (d) and (e), respectively. We can clearly observe the peak indicated by the red arrow, which corresponds to the three dynamic feature values of the tgs2610 at the  gas mixture of 180 ppm CO and 180 ppm $\rm H_2$ signal. Therefore, following the process of feature selection, a  total of 40 signal features were acquired for each trial.\par     
To observe the phenomenon of signal drift caused by the sensor array over time. The E-nose signal data from batch 11 (collected in June 2020), batch 12 (collected in June 2021), batch 13 (collected in March 2022), and batch 14 (collected in December 2022) were reduced from 40 dimensions to two dimensions using t-SNE. The resulting scatter plots are illustrated in Fig. \ref{fused_feature_compared2} (a). The global data distribution clearly reveals variability across batches 11, 12, 13, and 14. Furthermore, the local distribution of different gas types exhibits significant differences among these batches. Consistent with the experimental settings  outlined for the UCI data, the model was experimentally validated with batches 11 and 12 serving as the source domains, and batches 13 and 14 as the target domains, respectively. \par

The internal parameter settings for this experiment are indicated by the blue fonts in Fig. \ref{model}. The key parameters of the AMDS-PFFA model employed in the experimental setup are provided in Table \ref{para2}. All experiments were conducted with a consistent batch size of 16 and utilized a weight decay of $5\times10^{-4}$.  We also perform a comprehensive evaluation against the latest models with publicly available code. This comparison involved drift data collected from our self-developed E-nose system and multi-source domain models for related classification tasks. The single-domain approaches included DRCA, CDSL \cite{b42},  unsupervised feature adaptation (UFA) \cite{r27}, dynamic wavelet convolutional neural network (DWCNN) \cite{c2} integrated with the LMMD method, LDSP, dynamic wavelet coefficient map-axial attention network (DWCM-AAN) \cite{c3} integrated with the LMMD method,  hybrid attention-based transformer network with domain adversarial learning (HATN-DA) \cite{c4}, layer-normalized long short-term memory (LN-LSTM) \cite{c5} integrated with the LMMD method, and CDCNN,  while the multi-source domain methods compared were the multiple feature spaces adaptation network (MFSAN) \cite{c31}, moment matching for multi-source domain adaptation (M\textsuperscript{3}SDA) \cite{b43}, and semi-supervised domain adaptation (SSDA) \cite{c7}. The single-source domain approach considers both the optimal results from an individual source domain, denoted as ``single best" and those obtained by combining multiple source domains, denoted as ``source combine". The experimental gas recognition accuracy results are shown in Table \ref{acc_self}. Compared to recent gas sensor drift compensation methods reported in the literature, our proposed AMDS-PFFA model achieved the highest recognition accuracy: 95.09\% for target domain batch 13 and 92.92\% for target domain batch 14. Additionally, it recorded the highest average accuracy of 93.96\% across all methods tested. These results clearly demonstrate that our model provides superior resistance to sensor drift and outperforms the other models in the comparison.  \par 
\begin{table}[htbp]
	\centering
	\caption{Key parameters of ADMS-PFFA model employed in the experimental setup.}
	\scalebox{0.64}{
		\begin{tabular}{ccc}
			\toprule
			\toprule
			\textbf{Parameters} & \textbf{11, 12 \textrightarrow 13} & \textbf{11, 12 \textrightarrow 14} \\
			\midrule
			\textbf{Coefficient $\alpha$} & 1 & 5 \\
			\textbf{Learning rate} & 0.0018 & 0.006 \\
			\textbf{Batch size} & 16 & 16 \\
			\textbf{Weight decay} & $5\times10^{-4}$ & $5\times10^{-4}$ \\
			\textbf{Momentum} & 0.90 & 0.90 \\
			\textbf{Layers of S, G\textsubscript{1}, G\textsubscript{2}} & 3, 3, 3 & 3, 2, 2 \\
			\textbf{Dropout for S, G\textsubscript{1}, G\textsubscript{2}} & 0.2 & 0.2 \\
			\bottomrule
			\bottomrule
		\end{tabular}
	}
	\label{para2}
\end{table} 
To demonstrate the proposed model's effectiveness in reducing differences between source and target domain data distributions, thus mitigating gas sensor drift, the 2D distribution of fused features, reduced using the t-SNE method from source batches 11 and 12 across each target batch, is illustrated in Fig. \ref{fused_feature_compared2} (b) and (c). In Fig. \ref{fused_feature_compared2} (b), we can clearly observe that in pair 1, the overall data distribution between source batch 11 and target batch 13 is consistent, with well-concentrated and distinct local distributions of different gas types, displaying clear category separations. Similarly, in pair 2, the data distribution between source batch 12 and target batch 13 shows a consistent pattern, with concentrated local groupings and evident category distinctions. Fig. \ref{fused_feature_compared2} (c) further illustrates that in both pair 1 and pair 2, where batches 11 and 12 serve as the source domains and batch 14 as the target domain, the global and local data distributions consistently follow this pattern. These results confirm that the proposed AMDS-PFFA model not only effectively aligns the global distribution between source and target domains but also maintains clear distinctions in the local distribution of gas types, thereby mitigating sensor drift over time. \par 
\begin{table}[]
	\centering
	\caption{Comparison of gas identification accuracy (\%) for drift compensation models in the E-nose system.}
	\scalebox{0.60}{
		\begin{tabular}{lcccccc}
			\toprule
			\toprule
			\textbf{Standard} & \textbf{Model} & \textbf{Year} & \textbf{11, 12 --\textgreater{}13} & \textbf{11, 12 --\textgreater{}14} & \textbf{\begin{tabular}[c]{@{}c@{}}Average\\ accuracy\end{tabular}} \\
			\midrule
			\multirow{4}{*}{\textbf{\begin{tabular}[c]{@{}c@{}}Single \\ best\end{tabular}}}  
			& \textbf{DRCA} & 2017 & 80.99 & 79.75 & 80.34 \\ 
			& \textbf{CDSL} & 2017 & 75.81 & 64.88 & 70.10 \\ 
			& \textbf{UFA} & 2018 & 66.53 & 85.74 & 76.56 \\ 
			& \textbf{DWCNN+LMMD} & 2021 & 81.72 & 80.51  & 81.09 \\ 
			& \textbf{LDSP} & 2022 & 82.80 & 79.33 & 80.99 \\ 
			& \textbf{DWCM-AAN+LMMD} & 2023 & 90.54 & 87.99 & 89.21 \\ 
			& \textbf{HATN-DA} & 2023 & 85.59 & 82.28 & 83.86 \\ 
			& \textbf{LN-LSTM+LMMD} & 2024 & 81.51 & 78.15 & 79.76 \\ 
			& \textbf{CDCNN} & 2024 & 87.31 & 84.06 & 85.61 \\ 
			\midrule
			\multirow{3}{*}{\textbf{\begin{tabular}[c]{@{}c@{}}Source\\ combine\end{tabular}}} 
			& \textbf{DRCA} & 2017 & 75.03 & 82.64 & 79.00 \\ 
			& \textbf{CDSL} & 2017 & 91.94 & 65.50 & 78.14 \\ 
			& \textbf{UFA} & 2018 & 79.70 & 84.71 & 82.32 \\ 
			& \textbf{DWCNN+LMMD} & 2021 & 83.44 & 81.69 & 82.53 \\ 
			& \textbf{LDSP} & 2022 & 86.45 & 84.65 & 85.51 \\ 
			& \textbf{DWCM-AAN+LMMD} & 2023 & 92.26 & \underline{90.55} & \underline{91.37} \\ 
			& \textbf{HATN-DA} & 2023 & 87.96 & 81.30 & 81.48 \\ 
			& \textbf{LN-LSTM+LMMD} & 2024 & 84.52 & 80.51 & 82.43 \\ 
			& \textbf{CDCNN} & 2024 & 90.54 & 87.20 & 88.80 \\ 
			\midrule
			\textbf{Multi-source} 
			& \textbf{MFSAN} & 2019 & 87.50 & 79.17 & 83.15 \\ 
			& \textbf{M\textsuperscript{3}SDA} & 2019 & \underline{92.37} & 88.19 & 90.19 \\ 
			& \textbf{SSDA} & 2023 & 90.32 & 87.20 & 88.69 \\ 
			& \textbf{AMDS-PFFA} & 2024 & \textbf{95.09} & \textbf{92.92} & \textbf{93.96} \\ 
			\bottomrule
			\bottomrule
		\end{tabular}
	}
	\label{acc_self}
\end{table}
\vspace{-10pt} 
\subsection{Convergence study of the AMDS-PFFA model}  
\begin{figure}[htbp]\centering
	\includegraphics[width=8.8cm]{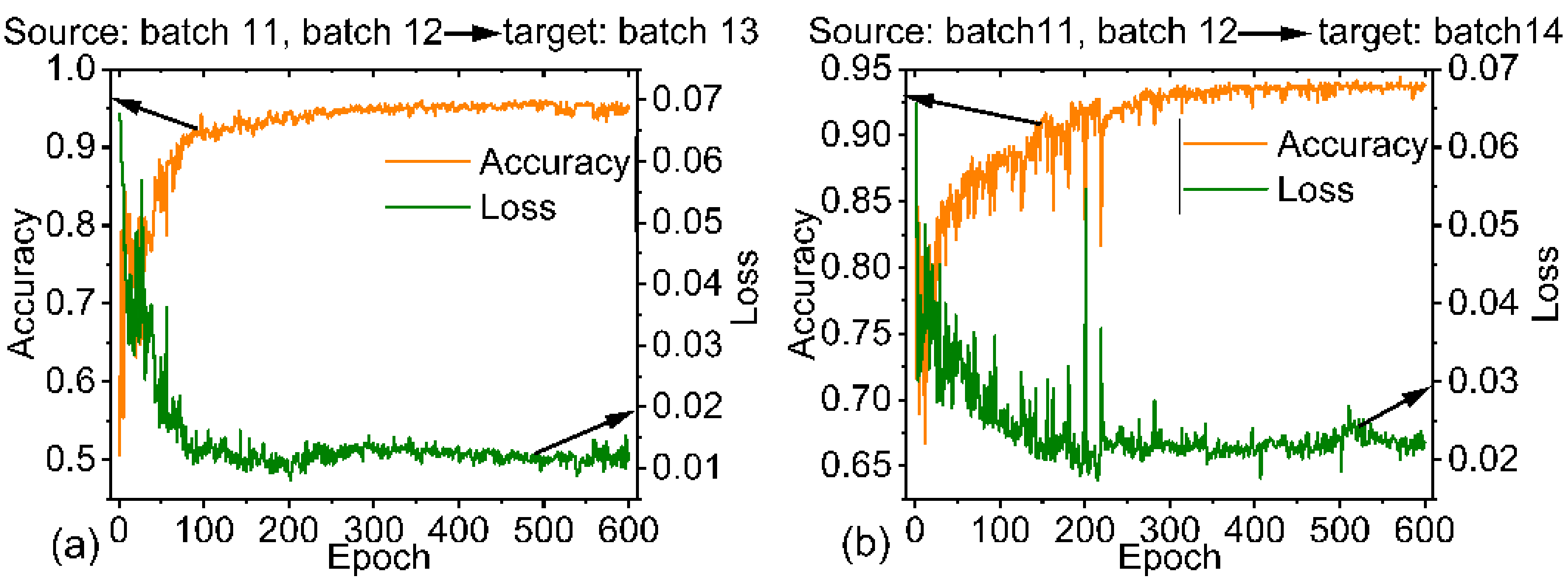}
	\caption{Convergence analysis of the unsupervised AMDS-PFFA model: accuracy and loss curves over epochs. (a) Target domain batch 13. (b) Target domain batch 14.}
	\label{acc_loss_batch1314}
\end{figure}

To evaluate the convergence of the proposed unsupervised AMDS-PFFA model, Fig. \ref{acc_loss_batch1314} (a) and (b) display the variations in loss and gas identification accuracy as the number of epochs increases. Both curves clearly show that the model converges consistently after approximately 250 epochs. Although minor fluctuations in accuracy and loss are observed during training, the curves stabilize over time. In particular, when using labeled feature data from source domains 11 and 12 to predict gas composition in the unlabeled target domain 14, notable fluctuations occurred around 200 epochs. However, by 250 epochs, the model exhibited strong convergence, reaching a stable state.\par

In summary, the proposed unsupervised AMDS-PFFA model effectively leverages labeled data from two source domains to enhance gas detection capabilities in the target domain while mitigating drift. The model has been experimentally validated using sensor array drift data from our self-developed E-nose system. The results demonstrate that AMDS-PFFA exhibits robust convergence and effectively mitigates drift in real-world E-nose applications, outperforming current drift compensation algorithms reported in the literature.

\section{Conclusion}
This study presents an unsupervised AMDS-PFFA model for drift compensation in gas identification using an E-nose system. By leveraging labeled feature data from multiple source domains, the model improves gas identification accuracy and compensates for drift in unlabeled target domain data. It addresses the challenges of long-term data annotation required by semi-supervised and supervised models, while also overcoming the limitations of existing domain adaptation methods that often neglect private, domain-specific features crucial for enhanced performance. The model's effectiveness is validated with drift data from both the UCI Chemosignals Lab and a self-developed E-nose system. Compared to recent drift compensation models, our model achieves the highest accuracies: 99.05\% for batch 3, 88.46\% for batch 4, 98.44\% for batch 5, 93.31\% for batch 6, 86.17\% for batch 7, 89.24\% for batch 8, and 66.83\% for batch 10 in the UCI dataset, with an overall average accuracy of 83.20\%. In the self-developed E-nose system, it reaches top accuracies of 95.09\% for batch 13 and 92.92\% for batch 14, with an average accuracy of 93.96\%. The model also exhibits strong convergence across both datasets.\par
Despite these achievements, the scope of the experiments may be limited, and the gas sensor data could be affected by various types of noise and interference. While the model is designed to address drift issues, its ability to handle noise and maintain data quality across different conditions remains a concern. Future work will focus on enhancing the model's robustness in the presence of diverse data quality challenges and expanding the experimental scope to include a broader range of real-world scenarios.

\end{document}